\documentclass[aps,prl,superscriptaddress,twocolumn,10pt,nofootinbib]{revtex4}
\usepackage[utf8]{inputenc}
\usepackage[english]{babel}
\usepackage{times}
\usepackage{color}
\usepackage{xcolor}
\usepackage{graphicx}
\usepackage{import}
\usepackage{amsmath}
\usepackage{braket}
\usepackage{algorithm}
\usepackage{algpseudocode}
\usepackage{tikz}
\usepackage{hyperref}
\usetikzlibrary{quantikz}
\usepackage{amsmath}

\makeatletter
\renewcommand*\env@matrix[1][*\c@MaxMatrixCols c]{%
  \hskip -\arraycolsep
  \let\@ifnextchar\new@ifnextchar
  \array{#1}}
\makeatother

\usepackage{booktabs}
\AtBeginDocument{
\heavyrulewidth=.08em
\lightrulewidth=.05em
\cmidrulewidth=.03em
\belowrulesep=.65ex
\belowbottomsep=0pt
\aboverulesep=.4ex
\abovetopsep=0pt
\cmidrulesep=\doublerulesep{}
\cmidrulekern=.5em
\defaultaddspace=.5em
}

\usepackage{tikz}
\usepackage{tikzscale}
\usetikzlibrary{backgrounds,fit,decorations.pathreplacing,calc,shapes}

\makeatletter
\def\BState{\State\hskip-\ALG@thistlm}
\makeatother

\begin{document}

\title{Transversal Injection: A method for direct encoding of ancilla states for non-Clifford gates using stabiliser codes.}

\author{Jason Gavriel}
\email{jason@gavriel.au}
\affiliation{Center for Quantum Software and Information, University of Technology Sydney.  Sydney, NSW, 2007, Australia.}
\affiliation{Centre for Quantum Computation and Communication Technology.}
\author{Daniel Herr}
\affiliation{d-fine GmbH, An der Hauptwache 7, 60213, Frankfurt, Germany.}
\author{Alexis Shaw}
\affiliation{Center for Quantum Software and Information, University of Technology Sydney.  Sydney, NSW, 2007, Australia.}
\affiliation{Centre for Quantum Computation and Communication Technology.}
\author{Michael J. Bremner}
\affiliation{Center for Quantum Software and Information, University of Technology Sydney.  Sydney, NSW, 2007, Australia.}
\affiliation{Centre for Quantum Computation and Communication Technology.}
\author{Alexandru Paler}
\affiliation{Aalto University, 02150 Espoo, Finland.}
\author{Simon J. Devitt}  
\affiliation{Center for Quantum Software and Information, University of Technology Sydney.  Sydney, NSW, 2007, Australia.}

\begin{abstract}
Fault-tolerant, error-corrected quantum computation is commonly acknowledged to be crucial to the realisation of large-scale quantum algorithms that could lead to extremely impactful scientific or commercial results.  Achieving a universal set of quantum gate operations in a fault-tolerant error-corrected framework suffers from a `conservation of unpleasantness'.  In general, no matter what error correction technique is employed, there is always one element of a universal gate set that carries a significant resource overhead - either in physical qubits, computational time, or both.  Specifically, this is due to the application of non-Clifford gates.  A common method for realising these gates for stabiliser codes such as the surface code is a combination of three protocols: state injection, distillation and gate teleportation.  These protocols contribute to the resource overhead compared to logical operations such as a CNOT gate and contribute to the qubit resources for any error-corrected quantum algorithm.  In this paper, we introduce a very simple protocol to potentially reduce this overhead for non-Clifford gates: Transversal Injection.  Transversal injection modifies the initial physical states of all data qubits in a stabiliser code before standard encoding and results in the direct preparation of a large class of single qubit states, including resource states for non-Clifford logic gates. Preliminary results hint at high quality fidelities at larger distances and motivate further research on this technique. 
\end{abstract}

\maketitle

Quantum Error Correction (QEC) forms a crucial component of large-scale quantum computing systems \cite{Devitt:2013aa,SD-Terhal:2015aa}.  The difficulty in fabricating ultra-low error rate qubits and quantum gates at scale necessitates active techniques to mitigate errors caused by environmental decoherence, fabrication errors, measurement and control errors, and components that are always probabilistic \cite{Rudolph:2017aa}.  While there is currently a focus on the so called NISQ regime \cite{SD-Preskill2018quantumcomputingin} - where it is hoped that a scientifically or commercially valuable quantum algorithm can be found that is small enough to not require QEC on the current or next generation quantum computing chipsets - most theoretical work suggests that the true value in quantum computing will lie with large simulation algorithms that will unarguably require extensive error correction \cite{Reiher:2017aa,PhysRevX.8.041015,PhysRevResearch.3.033055}, unless we see a significant revolution in hardware technology.  
\\
\\
While work on QEC is extensive \cite{Lidar:2013ts}, the physical constraints on quantum hardware architectures has resulted in the dominance of one type of QEC code, namely the surface code \cite{SD-Fowler:2012aa}.  Defined over a 2D nearest neighbour array of physical qubits, it is now the most studied QEC code and the preferred model for numerous architecture blueprints in multiple hardware platforms \cite{Lekitsch:2017aa,SD-Jones:2012aa,Mukai:2020aa,SD-Hill:2015aa,Bombin21}.
\\
\\
However, the implementation of QEC for {\em any} quantum algorithm, large or small, comes with a significant overhead in physical qubits and/or computational time \cite{Gidney2021}.  This is not surprising as the goal of QEC is often to take a physical error rate of the hardware of between $p = 10^{-3}\rightarrow 10^{-4}$ and reduce it by many orders of magnitude, with large-scale quantum simulation estimated to require error rates of $10^{-20}$ or even lower \cite{PhysRevX.8.041015,Matteo:2020aa}.   
\\
\\
Theoretical work in QEC, algorithmic design, compilation and resource optimisation has done a surprising job of figuring out better and better ways to implement error corrected algorithms \cite{Litinski2019gameofsurfacecodes,Fowler2018,GidneyFowler2019}, with one of the most studied algorithms, Shor's algorithm, a useful example.  Early compilation efforts, with the surface code, bench-marked Shor's algorithm at the beginning of the 2010's, showing that upwards of 30 billion components would be required to implement Shor-2048 \cite{Devitt:2013ab}.  By focusing entirely on better ways to implement both the algorithm itself and the underlying QEC protocols, this has been reduced to 20 Million qubits by the end of the 2010's without changing any assumptions at the physical hardware level \cite{Gidney2021}.  
\\
\\
How much this can still be reduced depends on several factors - even when we still do not change the hardware assumptions of the underlying microarchitecture.  The first is just the raw qubit overhead to encode a logical qubit of information up to some desired logical error rate.  For a distance $d$ error correction code, the logical error rate scales as $p_L \approx O(p^{\lfloor\frac{d-1}{2}\rfloor})$, under a simple symmetric Pauli model.  This assumes that the physical error rate of all parts of the hardware system (decoherence, control, measurement etc...) is under the fault-tolerant threshold of the code, approximately $p \approx 0.67\%$ for the surface code \cite{SD-Stephens:2014aa}. If we take a square, un-rotated, planar surface code, the total number of qubits (data + syndrome qubits) scales as $N = (2d-1)^2$, hence $p_L \approx O(p^{\lfloor\frac{\sqrt{N}-1}{4}\rfloor})$.  This exponential scaling means that for a heavily error corrected code, a lattice of $N > 1000$ is required\footnote{Multiple studies have numerically derived the precise scaling, including constant factors for a required logical error rate in the planar surface code as a function of additional hardware constraints, more precise physical error models and overall physical error rate \cite{SD-Stephens:2014aa}}.  How much this base level logical qubit overhead can be reduced, while still having a code that is architecturally feasible is still an open question. 
\\
\\  
This work introduces a simple new way to produce encoded non-Pauli Eigenstates.  This process we dub `Transversal Injection' modifies the way in which non-Clifford ancillary states are encoded. A standard approach for creating logical qubits is to initialise data qubits into the $\ket{0}$ or $\ket{+}$ states and measuring the stabilisers of the code. If all stabilisers commute or anti-commute in the desired fashion, we have successfully encoded a logical qubit in the $\ket{0}$ or $\ket{+}$ state respectively. Transversal injection involves performing a transversal rotation on all data qubits - initialising them in some non-Pauli state - and then following the same stabiliser measurement procedure. During the encoding state, the stabilisers will either commute or anti-commute and the encoded logical state will now be some non-Pauli eigenstate. The string of all stabiliser measurements forms what we will call a {\em stabiliser trajectory}, and can be used to determine the resultant state.

In this paper we perform simulations of this protocol under the influence of physical errors and investigate how the encoded error rates are related under a standard Pauli noise model on the circuit level. A modest post-selection strategy is also applied to improve the fidelity of this protocol. We show through these preliminary results that we can generate states with a lower fidelity than our physical error rate which eases the resource overhead of state distillation. Further research is needed to investigate fidelities at higher distances codes, further optimisation of the classical algorithm and compilation strategies. 
\\
\\
We present this new technique in the context of the surface code, but it should be stressed that it is applicable to all stabiliser based QEC codes. The contents of this paper includes:
\begin{itemize}
    \item \hyperref[link1]{A description of the formalism.}
    \item \hyperref[link2]{A classical algorithm for calculating the logically encoded state.}
    \item \hyperref[link3]{Numerical simulations of Transversal Injection on the surface code.}
    \item \hyperref[link4]{Discussion of this technique and implications for QEC circuit compilation.}
\end{itemize}
This new method can be looked at as the qubit extension of what was found in the continuous variable context \cite{Baragiola:2019tb}, where all-Gaussian universality was discovered in the context of the GKP code.

\section{State distillation}
Arguably the largest source of overhead in fault-tolerance is ancillary protocols required to perform error-corrected logic operations such as lattice surgery \cite{SD-Horsman:2012aa} and complex compound protocols to create resource states for universal computation \cite{Brav05,Litinski2019gameofsurfacecodes,litinski19}. Some gates have a transversal realisation in the code which is inherently fault-tolerant and require few if any additional resources to implement. Eastin and Knill \cite{East09} proved that there is no QEC code that can perform a universal set of transversal gates while correcting an arbitrary error. The no-go of Eastin and Knill can be worked around when not restricted to using a single code, such as Gauge-fixed code conversion \cite{SD-Bombin:2015aa}. A more common approach, is simply to `hack' together a fault-tolerant implementation of the non-transversal gate via a sequence of {\em state injection}, {\em magic state distillation} and {\em gate teleportation} \cite{Brav05,Bravyi:2012ul}. One example of state distillation is the quantum Reed-Muller code which takes 15 magic states with a respective error rate of $p$ and distils them into a single state with an error rate of $35p^3$. This process can be performed recursively to reach a desired level of fidelity.
\\
\\
The encoding of a magic state for use in state distillation has a $p$ that is dependent on the fidelity of the single qubit state, and the two-qubit gates that are needed to realise the encoding. Errors propagate in this protocol and causes the logically encoded state to have approximately the same {\em logical} error rate as the physical qubit and gates used for the injection, irrespective of the amount of error-correction used for the encoding~\cite{Li_2015}. Even small improvements at this stage of the protocol will be compounded by multiple state distillation steps, achieving a significant improvement in fidelity or a reduction in the amount of distillation needed for a target fidelity level. Any gain in fidelity seen through transversal injection could outweigh any complexity introduced by its probabilistic nature elsewhere in algorithm compilation.
\\
\\

\section{Pauli eigenstate encoding}
The actual procedure of transversal injection is only a minor alteration to standard surface code operation and is derived from a simple observation about how Pauli eigenstates are encoded. This section will briefly cover how a single high-fidelity logical state can be encoded using multiple physical qubits. This is a requisite for understanding how transversal injection works as described in the next section. 
\\
\\
Traditionally, we only consider two logically encoded states in the surface code that can be initialised, the $|0\rangle_L$ state and the $|+\rangle_L$ state.  These two states are eigenstates of the logical $Z_L$ and $X_L$ operators of the planar surface code and are hence natural to consider when examining encoded state initialisation.  The encoding procedure for each of these two states proceeds in a similar manner, we will use the $|0\rangle_L$ state for ease of discussion.  
\\
\\
When a logical qubit in the planar surface code is initialised into $|0\rangle_L$, we first initialise each of the physical qubits in the data block in the $|0\rangle$ state.  Hence our system can be described in terms of the following stabiliser matrix.
\begin{equation}
\begin{bmatrix} 
Z_1& I_2 & I_3 & ... & I_N \\
I_1 & Z_2 & I_3 & ... & I_N \\
I_1 & I_2 & Z_3 & ... & I_N \\
. & . & . & ... & . \\
I_1 & I_2 & I_3 & ... & Z_N 
\end{bmatrix}
\label{eq:init}
\end{equation}
where $N$ is the number of data qubits in the code, and the eigenvalue of each stabiliser is $+1$.  
\\
\\
Initialisation of the encoded state then requires repeated measurement of each of the $X$-type vertex stabilisers of the surface code \cite{SD-Fowler:2012aa}.  Each of these stabiliser measurements result in a random parity and the new stabiliser set will contain these $X$-type vertex stabilisers and combinations of the stabilisers in Eq.~\ref{eq:init} that {\em commute} with all the $X$-type stabilisers that were just measured. 
\\
\\
By definition, there are only two sets of operators that commute with all the $X$-type stabilisers, namely all the $Z$-type plaquette stabilisers of the code {\em and} the $Z$-chain operator that defines the $Z$-eigenstate of the encoded qubit.  The parity of these commuting operators are defined by the parity of the individual stabilisers of the physical $|0\rangle$ states that are the rows of Eq.~\ref{eq:init}.  
\\
\\
Hence after the $X$-type vertex stabilisers are measured (and their resultant syndromes decoded to perform the required $Z$-correction to transform the eigenvalues of the $X$ stabilisers into all +1), the stabiliser matrix will be
\begin{equation}
\begin{bmatrix}
X_{v_i} \\
.\\
Z_{p_i}\\
Z_L
\end{bmatrix}
\label{eq:init2}
\end{equation}
where $\{X_{v_i},Z_{p_i}\}$ for $i \in [1,(N-1)/2]$ are the stabilisers of the surface code and $Z_L$ is the chain operator that runs across the lattice, defining the logical state. 
\\
\\
The fact that each physical qubit starts in a $+1$ eigenstate of the $Z$ operator means that not only is it unnecessary to measure the $Z$-type plaquette stabilisers when initialising an encoded qubit, but that the resulting logical state will automatically be in a $+1$ eigenstate of $Z_L$. 
While in principle the $Z$-stabilisers do not need to be measured, in practice they are as once the qubit is initialised, it needs to be corrected against possible physical $X$-errors. 
\\
\\
As the parity of the resultant $Z$-stabilisers in Eq.~\ref{eq:init2} are determined by the product of the individual parities of the $Z$-stabilisers in Eq. \ref{eq:init}, you can derive what you would need to initialise a $|1\rangle_L$ state directly.  In the case of the surface code, initialising in the $|1\rangle_L$ requires initialising a subset of the physical qubits (e.g. along a {\em diagonal} of the lattice) in the $|1\rangle$ state, not {\em all} the physical qubits (which would actually initialise the logical qubit back into the $|0\rangle_L$ state).
\\
\\
\section{Transversal Injection}
\label{link1}
Transversal injection is the realisation that the traditional standard procedure is only a small subset of what is actually possible. For example, when starting with all physical qubits in the $|0\rangle$ state, encoding a $|0\rangle_L$ state can be done reliably as we are already in an eigenstate of the $Z$ portion of our stabiliser group. If instead our data qubits are all initialised in some arbitrary state $\ket{\chi}$, we don't simply encode our logical qubit into the same state as our physical states. The stabiliser group will now commute or anti-commute in a probabilistic fashion and the measured eigenvalues will infer what our encoded logical state is. 
\\
\\
\subsection{Initialisation}
Let's consider the case where all the data qubits are initially placed into the states
\begin{equation}
|\chi\rangle = \alpha|0\rangle + \beta|1\rangle
\end{equation}
i.e. we first perform a transversal rotation on all data qubits to the state $|\chi\rangle$ before we follow the same procedure to encode a $|0\rangle_L$ or $|+\rangle_L$ state.  
\\
\\
When considering the system of N data qubits that have been uniformly transformed, the values of each eigenstate are now naturally related to the hamming weight of its integer value and the chosen transversal operation. 
\begin{equation}
\ket{\chi}^{\otimes N} = \sum^{2^N-1}_{n = 0} \alpha^{N-\hat{H}(n)} \beta^{\hat{H}(n)} \ket{n}
\label{eq:ti_init}
\end{equation}
\\
\\
Where $n$ is the integer representation of the system's eigenvalues and $\hat{H}(n)$ is the Hamming weight, or number of bits set to 1 in its binary representation. 
\\
\\
\subsection{Syndrome Extraction}
The set of measurements for each stabiliser $\{X_{v_i}, Z_{p_i}\}$ defines a {\em stabiliser trajectory}. This set is denoted by the string $\{x_1,...,x_{(N-1)/2}, z_1,...,z_{(N-1)/2}\}$ of eigenvalues, $x_j, z_j \in \{+1,-1\}$, that are measured for the $N-1$ $X$-type vertex and $Z$-type plaquette stabilisers. This definition will be used throughout this paper to reference a set of measurement outcomes that together infer a resultant logical state. In the absence of physical errors, the $X$ projections are probabilistic and the $Z$ projections are deterministic when encoding a $|0\rangle_L$ or vice versa for the $|+\rangle_L$ state. In transversal injection the outcome of {\em all} measurements are probabilistic and will select out only certain eigenvalues that are consistent with the previously observed measurements.
\\
\\
The first step in encoding is to measure all the $X$-type stabilisers, $X_{v_i}$. Our initial state is projected into eigenstates of the $X$-stabilisers with the eigenvalue for each operator determined by the qubit measurements of each stabiliser circuit. The measurement of these stabilisers will result in an eigenvalue sequence for the $X$-type stabilisers and cause perturbations in the superposition of the $Z$-type stabilisers until they too are measured. The second step is to project into eigenstates of the $Z$-stabilisers in the same manner. 
\\
\\
The stabiliser tableau is defined below with the first and second $\frac{N-1}{2}$ rows referring to the $X$-type and $Z$-type stabiliser sets respectively. Similarly, the first and second $N$ columns are binary values indicating an $X$ or $Z$ Pauli operator respectively and the final column indicates the eigenvalue. By definition for the planar surface code, the $X$-type stabilisers only contain $X$ Pauli elements and vice versa for the $Z$ sector. The last row is the chain operator, spanning the lattice, that denotes the logical $Z$-operator.
\\
\\
\begin{equation}
\label{tableau}
\begin{bmatrix} [ccc|ccc|c]
x_{1,1} & \dots & x_{1,N} & z_{1,1} & \dots & z_{1,N} & X_{v_1} \\
\vdots & \ddots & \vdots & \vdots & \ddots & \vdots & \vdots \\
x_{\frac{N-1}{2},1} & \dots & x_{\frac{N-1}{2},N} & z_{\frac{N-1}{2},1} & \dots & z_{\frac{N-1}{2},N} & X_{v_\frac{N-1}{2}} \\
\hline
x_{\frac{N+1}{2},1} & \dots & x_{\frac{N-1}{2},N} & z_{\frac{N-1}{2},1} & \dots & z_{\frac{N-1}{2},N} & Z_{p_1} \\
\vdots & \ddots & \vdots & \vdots & \ddots & \vdots & \vdots \\
x_{N-1,1} & \dots & x_{N-1,N} & z_{N-1,1} & \dots & z_{N-1,N} & Z_{p_\frac{N-1}{2}} \\
\hline
x_{L,1} & \dots & x_{L,N} & z_{L,1} & \dots & z_{L,N} & Z_{L} \\
\end{bmatrix}
\end{equation}

When we initialise into the $|0\rangle_L$ state, we would now already have a logically encoded state (as all the $Z$-type stabilisers {\em including} $Z_L$ should automatically be satisfied), however these stabilisers are still measured anyway for a fault-tolerant initialisation. In the case of transversal injection $Z_L$ is now in some superposition of states, determined by the stabilisers that either commute or anti-commute — our stabiliser trajectory.
\\
\\
The logical state that the block is initialised into is also determined as,
\begin{equation}
z_{\text{chain}} = \left(\sum_{i \in \text{left/right}}z_i\right) \mod 2. 
\end{equation}
Where left/right is all qubits of a chosen chain that runs from the left boundary to the lattice to the right (defining the logical $Z$ operator).  If $z_{\text{chain}} =0$ we have initialised the logical qubit into the $|0\rangle_L$ state and if $z_{\text{chain}} =1$, the $|1\rangle_L$ state. 
\\
\\
When all qubits are initialised into $\ket{\chi}^{\otimes N}$ before encoding, we now are not simply in $\pm 1$ eigenstates of $Z_{p_i}$. Instead we are in a superposition of the $+1$ and $-1$ eigenstates.
\\
\\
Each possible stabiliser trajectory (indexed here by $i$) will infer a logical state, $\ket{\Lambda}_i$ and will have the following form after encoding,
\begin{equation}
\ket{\Lambda}_i=\begin{bmatrix} [c|c]  H_{x_1} & X_{v_1} \\
\vdots & \vdots \\
H_{x_{(N-1)/2}} & X_{v_{(N-1)/2}} \\
\hline
H_{z_1} & Z_{p_1} \\
\vdots & \vdots \\
H_{z_{(N-1)/2}} & Z_{p_{(N-1)/2}} \\
H_{Z_{\text{chain}}} & Z_L  
\end{bmatrix}_i = [\langle x_i, z_i, L_i\rangle],
\end{equation}
where $H_{x}$, $H_{z}, H_{Z_{\text{chain}}}$ is a concatenated form of \ref{tableau}. $x_i$ and $z_i$ are the eigenvalue bit strings corresponding to the measurements of the stabilisers, and $L_i$ is the eigenvalue bit string formed from the original $\ket{\Lambda}_i$ in each term, summed modulo 2. 
\\
\\
Using the syndrome measurements, we know how the stabiliser sets transform from the original diagonal $Z$ form to the stabilisers and logical operators for the encoded planar surface code. To calculate the resultant logical state, we now treat each term individually and follow them through the encoding procedure. 
\\
\\

\subsection{Resultant state}
As we are no longer in eigenstates of the logical $Z$ operator of the surface code, we will be left with some non-trivial logically encoded state. The probabilistic nature of the $X$ and $Z$-type stabiliser measurements means that the resultant encoded state from transversal injection is {\em probabilistic}, but {\em heralded}.
\\
\\
\begin{figure}[ht!]
	\includegraphics[width=3.5cm]{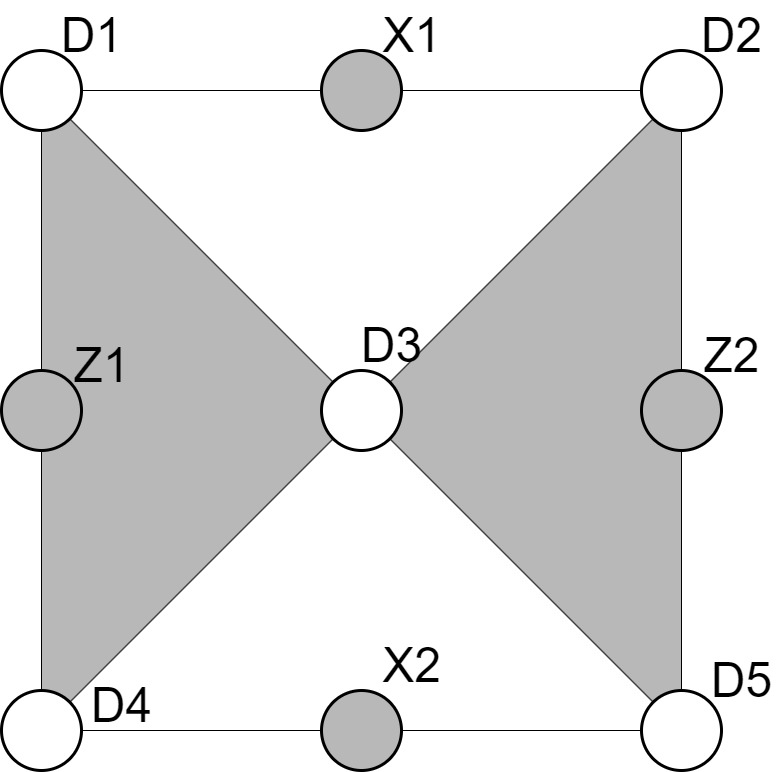}
	\caption{Qubit layout of the unrotated surface code at distance 2}
	\label{fig:d2_unrot_layout}
\end{figure}
The fact that we follow only one of the $2^{(N-1)}$ possible stabiliser trajectories when we initialise the code means we select out only the terms that are consistent with the eigenvalues we measure. For example, in the d=2 example in Fig. \ref{fig:d2_unrot_layout}, if $z_{p_1} = +1$, for stabiliser $Z_{p_1} = Z_{D1}Z_{D3}Z_{D4}$, then we can only keep the terms where $z_{p_1} = (z_{D1} + z_{D3} + z_{D4}) \mod 2 = 0$.  Any term where $z_{p_1} = 1$ will be inconsistent with the measurement projection that actually occurred and will be simply projected out of the resultant state.
\\
\\
What we are left with is a superposition of {\em only} the eigenvalues that are consistent with the $X$- and $Z$-stabiliser trajectories that are observed when initialising the state.  By definition - as we have now stabilised our data qubits with respect to {\em all} stabilisers of the planar surface code - we are going to be left with {\em some} superposition of the logic operator, $Z_L$, i.e. some new encoded state, $\ket{\Lambda}_i = \alpha_L|0\rangle_L + \beta_L|1\rangle_L$.  
Note that the resultant {\em logical} state does not, in general, match the transversal state, $|\chi\rangle$, that was used to initialise each of the physical qubits in the encoded block.  
\\
\\
To calculate the resultant logical state, we examine the last entry of the eigenvalue vector which is the logical observable, $L_i = z_{\text{chain}}$, formed from the eigenvalues of the individual data qubits that form a connected left/right chain through the planar surface code. The amplitude of the resulting $|0\rangle_L$ state, $\alpha_L$, will be the sum of all the terms where $L_i = 0$, while the amplitude of the $|1\rangle_L$, $\beta_L$ will be the sum of all the terms where $L_i=1$. After re-normalising the wave-function, we can now have an analytical form of the resultant encoded state as a function of $\alpha$, $\beta$ and $N$.  

\section{General algorithm for calculating the functional form of injected states}
\label{link2}
The above sections explain how analytical forms for the resultant logical state are calculated. We now want to present an explicit algorithmic construction that can be used to derive the logical state for an arbitrary distance, $d$, surface code.

\begin{algorithm}[H] 
\caption{Calculating a resultant logical state.}
\label{alg1}
\begin{algorithmic}[1]
\Require{$X_{1} \dots X_{(N-1)/2}$, $X$-stabilisers.} 
\Require{$Z_{1} \dots Z_{(N-1)/2}$, $Z$-stabilisers.} 
\Require{$L$, a chain operator for the logical $Z$-state.}
\Require{$\{\alpha$,$\beta\}$, coefficients of transversal physical states}
\Require{Number of data qubits of a distance $d$ planar surface code, un-rotated code, $N=d^2 + (d-1)^2$}

\Statex

\Function{Loop}{$\langle x_i \rangle \gets \hat{x}$} \# For each X-stabiliser trajectory
    \Function{Loop}{$n$ $\gets$ $1$ to $2^N-1$} \# For each eigenstate n
        \State $j = \text{Hamming}($n$)$.
        \State $\hat{k} = \ket{n} \times \langle x_i \rangle$ \# Project into eigenstates of X-stabilisers 
        \State $\lambda(\hat{k}) = j$ \# Dictionary of original Hamming weight
    \EndFunction
    \Function{Loop}{$\langle z_i \rangle \gets \hat{z}$} \# For each Z-stabiliser trajectory
        \State $\alpha_L(x_i, z_i) =$ $\beta_L(x_i, z_i) = 0$
        \Function{Loop}{$k_i$ $\gets$ $\hat{k}$} \# For each eigenstate n
            \State $m = \hat{k} \times \hat{z}$ \# Parity check with Z-stabilisers
            \State {\bf if $m \ne 0$}
                \State $j = \lambda(\hat{k})$
                \State $l = \hat{k} \times B$ \# Parity check with logical operator
            	\State {\bf if} $l = 0$, $\quad \alpha_L{(x_i, z_i)}$ = $\alpha_L{(x_i, z_i)} + \alpha^j \beta^{N-j}$.
            	\State {\bf if} $l = 1$, $\quad \beta_L{(x_i, z_i)}$ = $\beta_L{(x_i, z_i)} + \alpha^j \beta^{N-j}$.
	    \EndFunction
    \EndFunction
\State \Return {$\{\alpha_L({\hat{z}}), \beta_L({\hat{z}})\}/\sqrt{|\alpha_L(\hat{z})|^2+|\beta_L(\hat{z})|^2}$}
\EndFunction
\end{algorithmic}
\end{algorithm}
For example, for a distance $d=2$ code (Fig. \ref{fig:d2_unrot_layout}), we have a total of $N = 5$ physical qubits, each of the matrices, $M$, are $5\times 5$ matrices formed from the two $X$-stabilisers, two $Z$-stabilisers and one logical operator of the distance two surface code,
\begin{equation}
M = \begin{pmatrix}
1 & 1 & 1 & 0 & 0 \\
0 & 0 & 1 & 1 & 1 \\
1 & 0 & 1 & 1 & 0 \\
0 & 1 & 1 & 0 & 1 \\
1 & 1 & 0 & 0 & 0 
\end{pmatrix}.
\end{equation}
If we are to only consider the trivial $X$-stabiliser trajectory ($\ket{0}$ syndromes in $X$), we consequently find the four sets of analytical equations (up to re-normalisation) corresponding to the four sets of trajectories, $\hat{t} = \{0000,0001,0010,0011\}$.
\begin{equation}
\begin{aligned}
\ket{\Lambda}_{00} = &\begin{pmatrix} \alpha_L \\ \beta_L \end{pmatrix} =\begin{pmatrix} \alpha^5 + 2\alpha^2\beta^3 + \alpha\beta^4 \\ 2\alpha^3\beta^2+2\alpha^2\beta^3 \end{pmatrix} \\
\ket{\Lambda}_{11} = &\begin{pmatrix} \alpha_L \\ \beta_L \end{pmatrix} =\begin{pmatrix} \beta^5 + 2\beta^2\alpha^3 + \beta\alpha^4 \\ 2\beta^3\alpha^2+2\beta^2\alpha^3 \end{pmatrix} \\
\ket{\Lambda}_{01} = \hat{10}, &\begin{pmatrix} \alpha_L \\ \beta_L \end{pmatrix}  = (\alpha^4\beta + \alpha^3\beta^2+\alpha^2\beta^3 + \alpha\beta^4)\begin{pmatrix} 1 \\ 1\end{pmatrix} \\
\end{aligned}
\label{eq:final}
\end{equation}

For two (out of four) measured trajectories we get a non-trivial set of logical states ($\hat{t} = \{0000,0011\}$). For the other two stabiliser trajectories ($\hat{t} = \{0001,0010\}$) we simply initialise into the logical $\ket{+}_L = \left(\ket{0}_L+\ket{1}_L\right)/\sqrt{2}$ state.

\begin{figure}[ht!]
    \includegraphics[width=0.70\columnwidth]{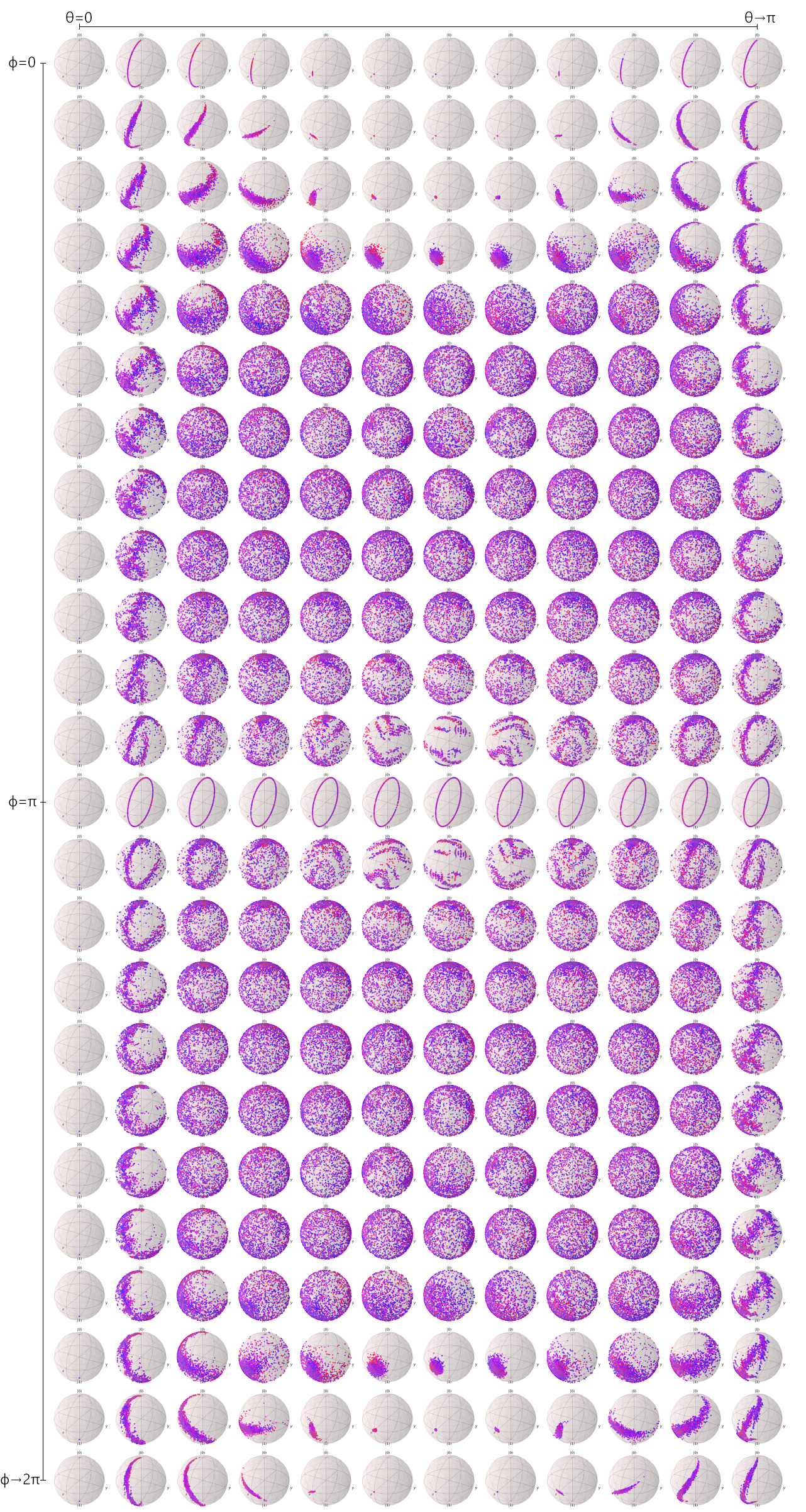}
	\caption{The possible logical states that result from transversal injection on a distance $d=4$ code for a range of initial $\theta$ and $\phi$ values. Each sphere is a plot for a specific initial rotation. Each dot on the sphere corresponds to a possible stabiliser trajectory and its respective output state.  Generated using QuTiP~\cite{johansson_qutip_2013}.}
	\label{fig:d4}
\end{figure}

This generalised algorithm will allow for the calculation of any transversely realisable encoded state for an arbitrarily large distance code, $d$: it intuitively follows the logic of the stabiliser execution.
\\
\\
Line 5 requires an explicit calculation of \emph{all} vectors generated from the $X$-stabiliser projections and will scale exponentially as a function of code distance, $d$. This proves to be a difficult task for classical computers as there are an exponential number of trajectories to track during the execution of the algorithm. Additionally, the solution space is large and storing the final results would be infeasible for large distances. 
\\
\\
There is an optimised version of this algorithm, which we introduce in the appendix. The optimised algorithm is based on a different approach and instead tracks only the terms relevant to a single target trajectory. We can now compute the logical state of this trajectory in a more efficient manner. Algorithm 2 scales exponentially with $(N-1)/2$, the number of $Z-$stabilisers, rather than with the number of physical qubits. Hence, calculating a target trajectory is $\mathcal{O}(e^{(N-1)/2})$ which is an improvement from $\mathcal{O}(e^{N})$ required to derive all possible trajectories. This improvement is most noticeable in the memory requirements which blow out in the first algorithm but can be dealt with comfortably in algorithm 2. Using GPU acceleration, we have demonstrated this algorithm on trajectories with trivial $X$-measurements on systems with over 100 data qubits. When non-trivial $X$ measurements are considered, classical computation of distance 5 is achievable using algorithm 2. This opens up the opportunity for a just-in-time strategy where output states from transversal injection can be calculated classically and then used as part of a broader compilation strategy. To realise this strategy, a more efficient algorithm will be needed for larger distances required for large-scale, error-corrected computation. 
\\
\\
In Fig. \ref{fig:d4}, we have plotted the distribution of resultant states across the Bloch sphere for a range of initial parameters. There are clear structures that form depending on the parameters of the initial rotation including great circles, arcs with different radii, clusters around poles and dense packing of the entire sphere. Certain distributions may be advantageous for certain compilation strategies or for physical systems that have error biases in certain directions. Alternatively, having a dense covering of the Bloch sphere might be optimal in a compilation strategy that benefits from a large set of distinct states.

\section{Realising non-Clifford logic operations through transversal injection}
The primary benefit of this new technique is producing logically encoded non-Pauli states with a higher fidelity directly onto the surface code. This aims to reduce the qubit and time resources required to distil magic states and in turn the quantity of costly non-Clifford logic gates. 

\begin{figure}[ht!]
	\includegraphics[width=0.9\columnwidth]{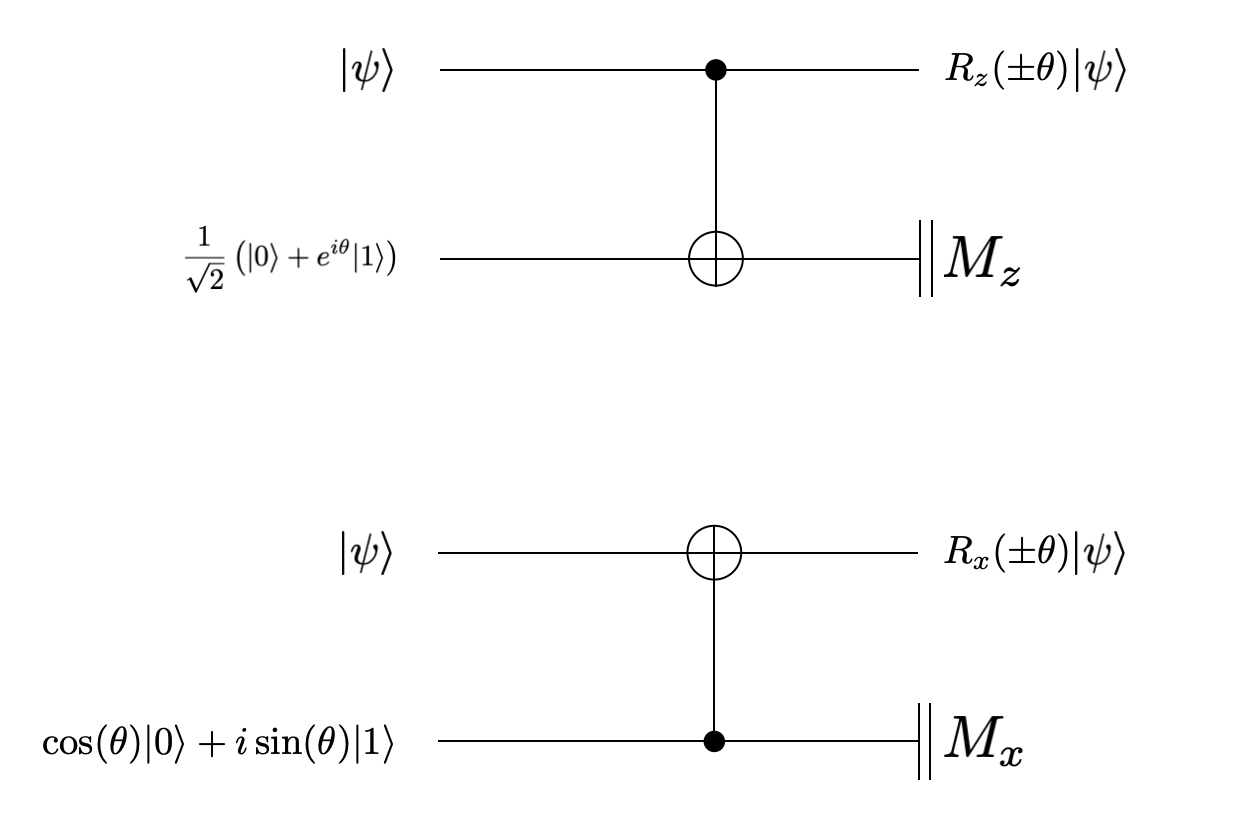}
	\caption{Quantum circuits to use non-Clifford states to enact non-Clifford gates.}
	\label{fig:teleportation}
\end{figure}

Consider the circuits shown in Fig. \ref{fig:teleportation}
In each of these circuits, we have our data qubit, $|\psi\rangle$ and an ancilla qubit that is prepared in the states, $\left(|0\rangle+e^{i\theta}|1\rangle\right)/\sqrt{2}$ or $\cos(\theta)|0\rangle + i\sin(\theta)|1\rangle$.  The rest of the circuit consists of the Clifford gates, CNOT, and measurements in either the $X$- or the $Z$-basis.  It can be easily verified that after the ancilla is measured, the resultant state is the rotated gate, $R_z(\pm \theta)$ or $R_x(\pm \theta)$ applied to the data qubit, $|\psi\rangle$.  The angle, $\theta$ is determined by the form of the ancillary state and the $\pm$ is determined by the outcome of the measurement result on the ancilla.  Consequently, the ability to perform arbitrary single qubit rotations becomes a problem of preparing appropriate ancillary states. 

To match up the functional forms in Eq. \ref{eq:final} to the functional forms necessary for the circuits in gate teleportation, in the context of our d=2 example, we need to solve the following.
\begin{equation}
\begin{aligned}
\begin{pmatrix}
1 \\ e^{i\theta} 
\end{pmatrix} = &\begin{pmatrix} \alpha^5 + 2\alpha^2\beta^3 + \alpha\beta^4 \\ 2\alpha^3\beta^2+2\alpha^2\beta^3 \end{pmatrix} \\ \text{or} &\begin{pmatrix} \beta^5 + 2\beta^2\alpha^3 + \beta\alpha^4 \\ 2\beta^3\alpha^2+2\beta^2\alpha^3 \end{pmatrix} \\
\text{and} \begin{pmatrix}
\cos(\theta) \\ i\sin(\theta)
\end{pmatrix} = &\begin{pmatrix} \alpha^5 + 2\alpha^2\beta^3 + \alpha\beta^4 \\ 2\alpha^3\beta^2+2\alpha^2\beta^3 \end{pmatrix} \\ \text{or} &\begin{pmatrix} \beta^5 + 2\beta^2\alpha^3 + \beta\alpha^4 \\ 2\beta^3\alpha^2+2\beta^2\alpha^3 \end{pmatrix} \\
\end{aligned}
\end{equation}
for $\alpha$, $\beta$ and $\theta$ (up to re-normalisation), in the case of the $d=2$ solution.  Once we find specific forms of $\alpha$ and $\beta$ that give rise to these functional forms, we know what transversal injected states are needed on the physical qubits to generate the correct form of the encoded states to be used in the circuits in Fig. \ref{fig:teleportation} to realise $R_z(\theta)$ and $R_x(\theta)$ rotational gates. 

\clearpage
\onecolumngrid

\clearpage
\twocolumngrid
\subsection{Eastin-Knill Theorem}
Transversal injection circumvents the Eastin-Knill no-go theorem as it is a method for preparing non-Clifford resources, such as a T-gate, to achieve universal quantum computation on the planar surface code. This new technique does not violate the Eastin-Knill theorem and is in fact still limited by the implications of the theorem. Non-Clifford gates must be realised in the fashion described in the preceding section, a T-gate still can't be implemented transversally on this code. Additionally the encoding step in this technique has a diminished error detecting ability compared to standard surface code operation. Some errors in the first round of stabiliser measurements can no longer be detected. Such errors will herald an incorrect stabiliser trajectory which infers that the system is in a different logical state than it actually is. Post-selection has the potential to mitigate this. However, we are still ultimately bound by this theorem.

\begin{figure}[ht!]
	\includegraphics[width=0.9\columnwidth]{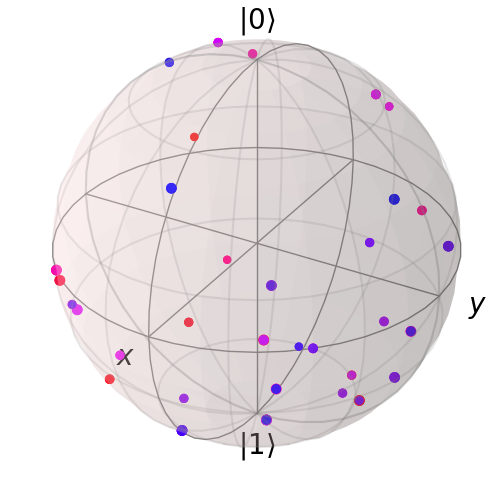}
	\caption{The 64 logical, single qubit, states (only trivial $X$-stabiliser measurements), resultant from transversal injection on a distance $d=3$ unrotated code from the table above. Generated using QuTiP~\cite{johansson_qutip_2013}.}
	\label{fig:d3}
\end{figure}

\section{Numerical simulations of error scaling}
\label{link3}
Typically an encoded state produced using the planar surface code will exhibit asymptotic fault-tolerance. Transversal injection does not introduce any new multi-qubit gate operations beyond what is required for standard $|0\rangle_L$ and $|+\rangle_L$ state initialisation so one may expect the same scaling out of this protocol. We confirm this assumption with numerical simulation where errors occur {\em only} after encoding, however this scenario is unrealistic and the protocol as a whole must be considered.
\\
\\
The simulations shown in this paper are performed with physical errors during all rounds of stabilisers, including the first stabiliser measurements. This results in an encoded error rate greater than the physical error rate $p$. Experimenting with post-selection strategies indicates that an error rate not bound by $p$ is in fact achievable, warranting further research into the behaviour of transversal injection at distance 5 and higher. 
\\
\\
To test transversal injection, surface code circuit simulations were performed using a balanced Pauli noise model. As we need to verify a non-trivial final state against the output from the derived logical state above, we used a full state simulator rather than stabiliser simulators such as Stim\cite{gidney2021stim}. The circuit simulations were performed with $d$ rounds of stabiliser measurements followed by a final perfect round of measurements. For all numerical data in this paper any syndrome is considered valid, however simulations where the syndromes change over multiple sweeps are discarded.
\\
\\
The unrotated surface code was used for the worked examples and for initial testing. However, the rotated surface code was subsequently chosen for numerical simulations as the lower number of qubits allowed a statistically significant amount of data to be acquired. Additionally, the numerical data that is presented in this paper was collected by re-using ancilla qubits for syndrome extraction. Since we are not constricted to nearest neighbour architecture in classical simulations, this allows us to profile transversal injection up to distance 4 across different uniform physical error levels and differing single/two-qubit error rates. 
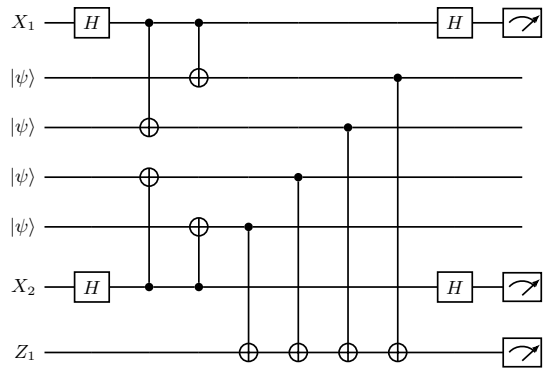
\begin{figure}[ht!]
\begin{tikzpicture}
\node[scale=0.8] {
\begin{quantikz}
\lstick{$X_1$}  & \gate{H} & \ctrl{2} & \ctrl{1} & \qw & \qw & \qw & \qw & \gate{H} & \meter{}\\
\lstick{$\ket{\psi}$} & \qw & \qw & \targ{} & \qw & \qw & \qw & \ctrl{5} & \qw & \qw\\
\lstick{$\ket{\psi}$} & \qw & \targ{} & \qw & \qw & \qw & \ctrl{4} & \qw & \qw & \qw\\
\lstick{$\ket{\psi}$} & \qw & \targ{} & \qw & \qw & \ctrl{3} & \qw & \qw & \qw & \qw\\
\lstick{$\ket{\psi}$} & \qw & \qw & \targ{} & \ctrl{2} & \qw & \qw & \qw & \qw & \qw\\
\lstick{$X_2$}  & \gate{H} & \ctrl{-2} & \ctrl{-1} & \qw & \qw &  \qw & \qw & \gate{H} & \meter{}\\
\lstick{$Z_1$}  & \qw & \qw & \qw &  \targ{} & \targ{} & \targ{} & \targ{} & \qw & \meter{}\\
\end{quantikz}
};
\end{tikzpicture}
\caption{Circuit diagram of the rotated surface code at distance 2.}
\label{fig:d2_circuit}
\end{figure}

\begin{figure}[ht!]
	\includegraphics[scale=0.12]{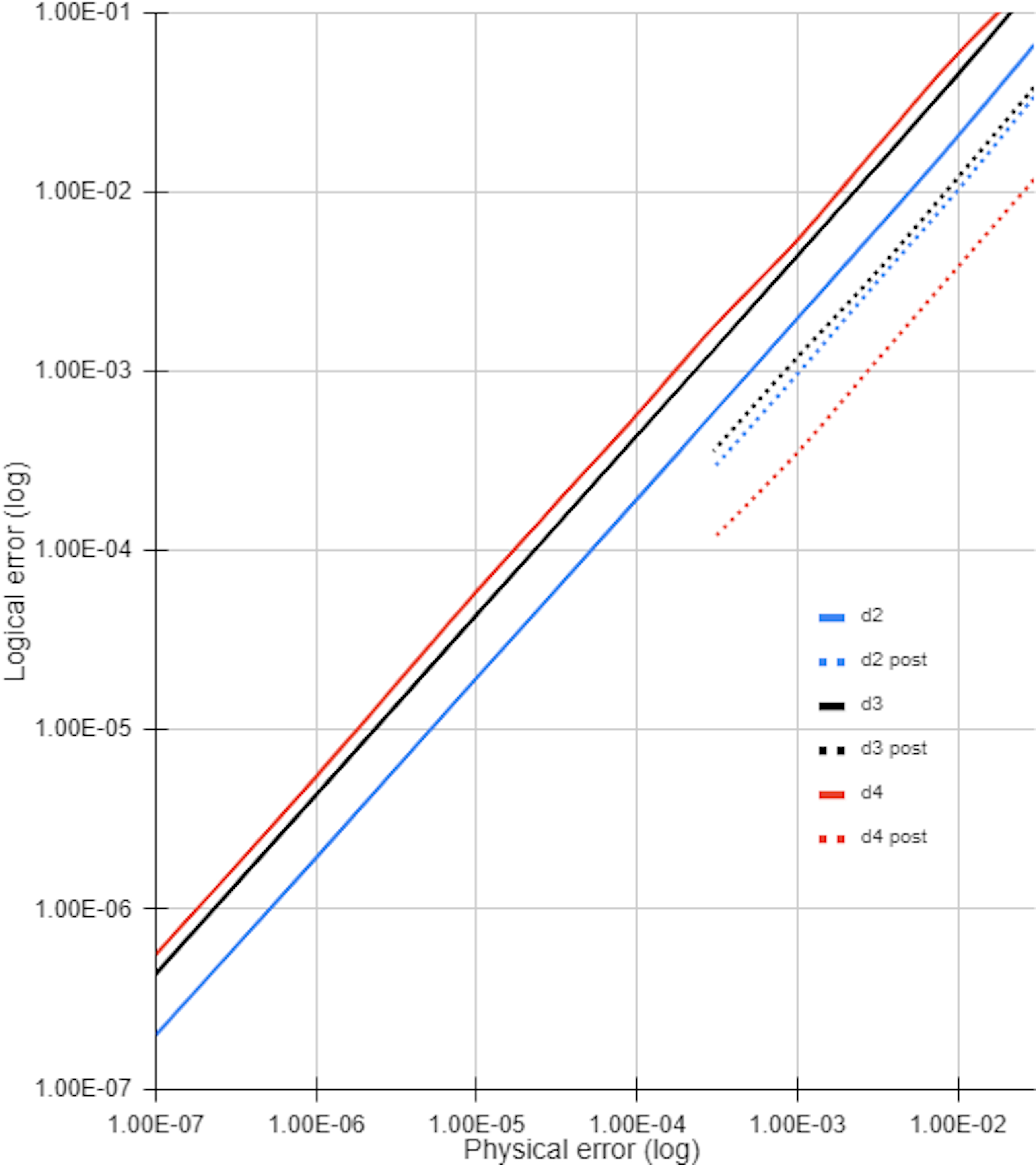}
	\caption{Scaling of logical fidelity vs. physical error rate. Fidelity is measured as the overlap between the simulated encoded state and the expected encoded state heralded by the stabiliser measurements of each run. $\theta_{initial} \approx 1.7728, \phi_{initial} \approx 3.3237$}
	\label{fig:results2}
\end{figure}
\begin{figure}[ht!]
	\includegraphics[scale=0.12]{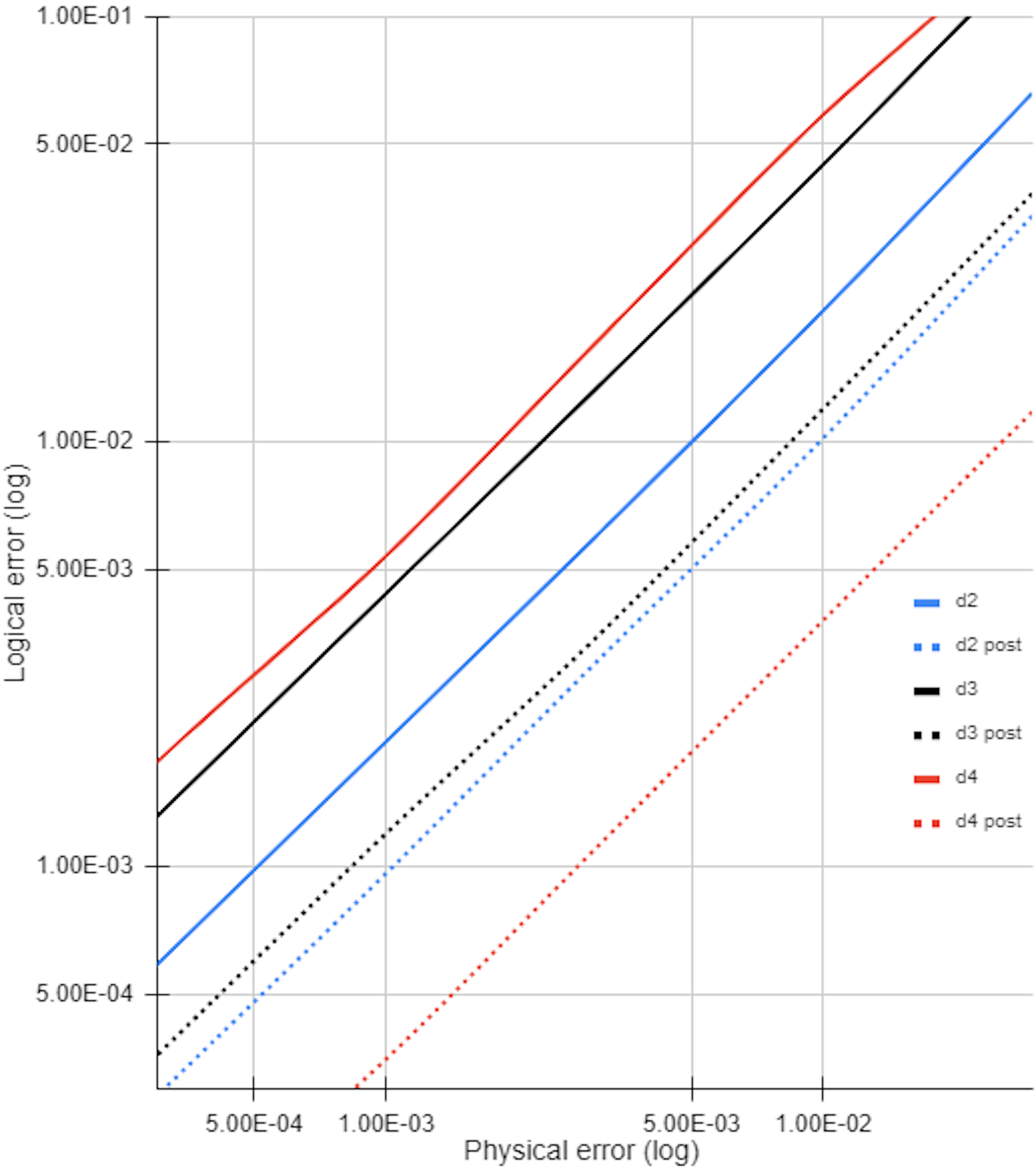}
	\caption{This figure is the same set of results as Fig. \ref{fig:results2}, inspecting a smaller range of physical error rates to focus on the post-selection results.  $\theta_i \approx 1.7728, \phi_i \approx 3.3237 $}
	\label{fig:results}
\end{figure}

\begin{figure}[ht!]
	\includegraphics[width=3.5cm]{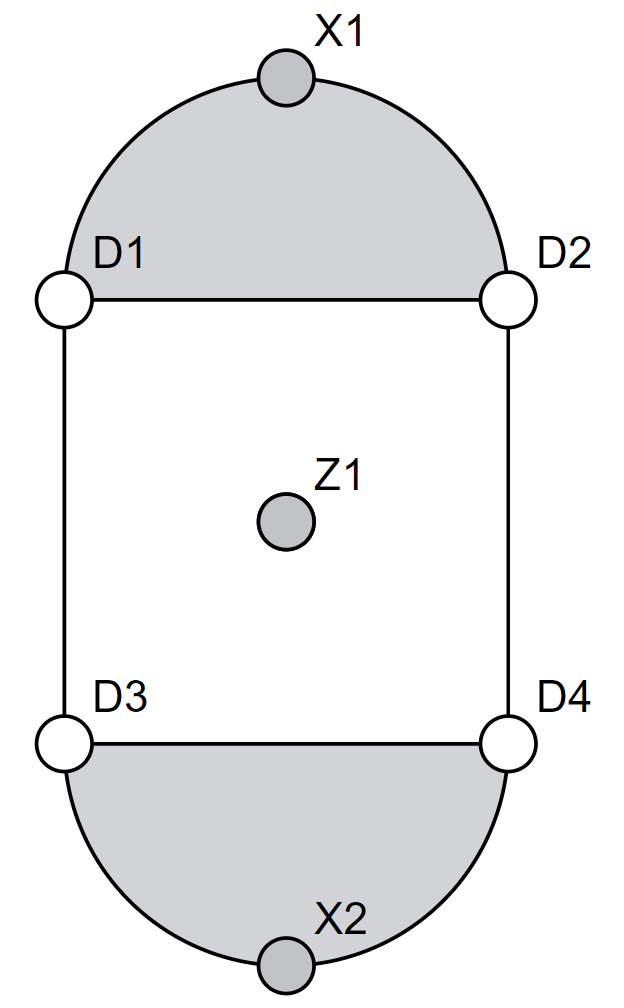}
	\caption{Qubit layout of the rotated surface code at distance 2.}
	\label{fig:d2_layout}
\end{figure}
First, a transversal rotation of all data qubits is performed. Ancillary qubits perform CNOT operations on their nearest neighbours in the $X$- and $Z$- basis for the vertex and plaquette stabilisers respectively. The ancillary qubits are measured, extracting the syndrome as is standard for the surface code. The syndrome measurements and parameters of the transversal rotation are passed into algo. \ref{alg2} where the state of our system is returned in an analytical form. The full state of our simulated system is yielded and compared to the result from algo. \ref{alg2} to determine if a logical error has occurred. 
\\
\\
Fidelity did appear to vary for different initial states and a chosen state too close to a pole on the Bloch sphere would 'snap' to that pole for a large portion of the trajectories. Hence, the amount of unique non-Clifford logical states seems to increase proportionately to how far away an initial rotation is from the poles. 
\\
\\
The relationship between physical error rate and logical rate are linearly proportional as we approach physical error rates below $p=0.01$ (Fig. \ref{fig:results2}). It appears that higher distance codes perform worse than lower distance codes and the logical error rate always exceeds the physical error rate. Without a post-selection strategy, numerical simulations can be done more efficiently at lower physical error rates.
\\
\\
For each distance and physical error rate, a second experiment was run where a post-selection strategy was employed in an attempt to improve fidelity (Fig. \ref{fig:results}). Post-selection does appear to yield improvements in fidelity by filtering out trajectories of simulated runs with a naive, pre-computed lookup table. For distance 2, this lookup table was only populated with the trivial syndrome as it is the only one that results in non-Clifford states. Distance 4 with post-selection appears to yield a significant improvement in fidelity, dropping {\em below} the physical error rate to roughly $0.39p$. All of the above experiments were benchmarked with non-uniform single qubit error rates of $p$, $0.1p$ and $0p$ with no measurable difference in fidelity.

To construct the lookup table used for an experiment, statistical analysis was done over a large number of simulations and the trajectories were ranked according to their average fidelity. A budget of $\approx 20\%$ was allocated and the top trajectories were selected until this budget was satisfied. An example for demonstration (not experimental data) is in Table. \ref{Tab:exampledata}. Given this data, we would whitelist the top performing trajectories (000, 011 and 101) until our $\approx 20\%$ quota was satisfied.

\begin{center}
\begin{table}[ht]
\caption{Example for demonstrating a post-selection strategy}
\begin{tabular}{ |c|c|c| } 
\hline
Trajectory & Fidelity & Frequency \\
\hline
011 & 99.99 & 0.01 \\ 
000 & 99.89 & 0.10 \\ 
101 & 98.58 & 0.09 \\ 
001 & 98.01 & 0.20 \\ 
100 & 97.84 & 0.31 \\ 
010 & 95.43 & 0.20 \\ 
110 & 95.21 & 0.05 \\ 
111 & 94.99 & 0.04 \\ 
\hline
\end{tabular}
\label{Tab:exampledata}
\end{table}
\end{center}

\section{Consequences for error-corrected circuit compilation}
\label{link4}
Transversal injection has the potential to reduce ancilla requirements and the amount of state distillation — however, it is not without its own drawbacks. Transversal injection does provide us with non-Clifford, logically encoded states that are fault tolerant after initialisation. The initialisation step itself is not a fault tolerant operation and is susceptible to two-qubit correlated errors. If these errors occur before the first stabiliser measurements are extracted, the initial state will be altered without detection events. Single qubit errors on ancillary qubits are either detected in subsequent syndrome extractions or change the logical state in a way that is heralded by its measurement (i.e. once the initial Pauli frame of an encoded state is known, the encoded state is defined).
\\
\\
Current methods of magic state encoding similarly exhibit a linear scaling between the encoded state's fidelity and physical error rate\cite{Li_2015}. Generally, the fidelity of output states from transversal injection is worse until a post-selection strategy is applied. Once post selection is applied on our distance 4 simulations, the logical error rate is comparable with the lower bound of other results even at much larger distances. To achieve the same fidelity, transversal injection has a much smaller qubit footprint and only requires a moderate amount of post-selection. Further analysis is needed to evaluate this protocol at distance 5 and higher to determine how fidelity scales beyond the results in this paper. 
\\
\\
Transversal injection is intrinsically a probabilistic method for encoded state preparation and while it is heralded, it is not something that is controllable.  Consequently when utilising transversal injection as a method for realising universality, special care must be taken when examining how to effectively compile error-corrected circuits.  
\\
\\
While a large number of states on the {\em logical Bloch Sphere} are available using this technique, the number of possible stabiliser trajectories scales exponentially as a function of the number of qubits and hence code distance.  In our preliminary analysis, the number of Pauli eigenstates that are prepared become exponentially unlikely as the code distance is scaled. There are potential redundancies at higher distances, but so far we have not identified any patterns between the exponential number of states that are possibly encoded for a fixed code distance.  
\\
\\
This implies that when a specific ancillary state is required for a teleported gate, it will need to be constructed through an effective random walk over the Bloch sphere. The exponential number of states on the Bloch Sphere that are available implies that rather than compiling to simply the non-Clifford $T$-gate in an error corrected system, we should be able to compile to any $Z$ or $X$ axis rotation we want by producing random logical states and approximating the required ancillary state needed for a given $R_z(\theta)$ or $R_x(\theta)$. Transversal injection can be used to produce magic states which we can then distil to an arbitrary accuracy, although this is only a subset of the possible output states. Hence, there is motivation to research distillation methods that are effective on other non-Pauli states. This has clear implications for circuit level compilation as the Clifford + $T$ alphabet for a fault-tolerant compatible circuit will no longer be a constraint.  
\\
\\
There are various techniques developed in the literature in approximating single qubit gates via a random walk around SU(2) that can be exploited to find a systematic solution to the gate compilation issue \cite{Brav05}, but this is relegated to further work. 
\\
\\
While a direct resource comparison to a compiled algorithm using magic state distillation, such as Shor-2048 \cite{Gidney2021} will require a systematic solution to compiling arbitrary single qubit logical rotations, there is a potential for reduced qubit requirements for any large-scale algorithm by utilising this new technique.

\section{Conclusions}
We have presented a new technique for achieving fault-tolerant universal quantum computation in a stabiliser code environment without changing any other operating assumption of the underlying code.
\\
\\
While we have presented this new scheme in the context of the surface code, transversal injection will be possible using any stabiliser based QEC code and the algorithm detailed in this work can be used to pre-compute resultant logical states, depending on the code structure and size.  
\\
\\
Interesting further work includes understanding if this technique can be used for codes beyond stabiliser codes and how transversal injection can be used in fully compiled error corrected algorithms.  Incorporating transversal injection into fully compiled, large-scale circuits, will allow us to re-benchmark algorithms such as Shor's algorithm or error corrected chemistry simulations to determine the exact resource savings over state of the art compiled results.
\\
\\
It should be noted that the algorithms presented in this work scale exponentially with the number of $Z-$stabilisers. Currently, we are able to compute explicit analytical forms for any stabiliser trajectory up to a $d=8$, non-rotated, surface code (consisting of $N = 113$ data qubits). This will be more than sufficient for any experimental demonstration in the near term. With further development and specialised hardware, higher distances will likely be possible to calculate.
\\
\\
However, we anticipate that a more efficient technique can be developed that allows for analytical forms to be calculated fast.  This may not allow for {\em all} of the exponential number of trajectories to be pre-calculated, but if a single trajectory for an arbitrary distance can be calculated fast, then this will be sufficient for a {\em just in time} approach to be taken, where analytical forms for logical states are calculated at the time they are physically created in an error corrected machine. 
\\
\\
This work is a potential solution to some of the overheads blocking the path to large-scale, error-corrected computation and provides another approach to circumvent the Eastin-Knill theorem, allowing for universal, error-corrected computation in an increasingly resource efficient manner.

\section{Acknowledgements}
Thank you to Austin Fowler for early feedback on the findings of this paper. The views, opinions and/or findings expressed are those of the authors and should not be interpreted as representing the official views or policies of the Department of Defense or the U.S. Government. This research was developed in part with funding from the Defense Advanced Research Projects Agency [under the Quantum Benchmark- ing (QB) program under award no. HR00112230007 and HR001121S0026 contracts]. MJB acknowledges the support of Google. MJB, JG, and AS, were supported by the ARC Centre of Excellence for Quantum Computation and Communication Technology (CQC2T), project number CE170100012. AS was als supported by the Sydney Quantum Academy.

\bibliographystyle{unsrt}
\bibliography{bib2}

\clearpage
\onecolumngrid
\appendix

\section{Appendix A - Optimisation of algorithm 1.}
We can provide this refinement to Algo. \ref{alg1} as detailed below. 
\begin{algorithm}[H] 
\caption{Calculating a specific logical state.}
\label{alg2}
\begin{algorithmic}[1]
\Require{$X_{1} \dots X_{(N-1)/2}$, $X$-stabilisers.} 
\Require{$Z_{1} \dots Z_{(N-1)/2}$, $Z$-stabilisers.} 
\Require{$L$, a chain operator for the logical $Z$-state.}
\Require{$\{\alpha$,$\beta\}$, coefficients of transversal physical states}
\Require{Number of data qubits of a distance $d$ planar surface code, un-rotated code, $N=d^2 + (d-1)^2$}
\Require{$\langle x_i \rangle, \langle z_i \rangle$ eigenvalue bit strings corresponding to stabiliser measurements}

\Statex
\State {$\hat{m} = \mathbf\{\}$ \# List of set stabiliser measurements $(N-1)/2$}
\State {$\hat{v} = [\{0\}_N]$} \#Eigenvalue memory, initially just $\{0\}_N$

\Function{Loop}{$i$ $\gets$ $1$ to $(N-1)/2$} \#For Each $Z$-stabiliser
    \State {$\hat{v_t} = []$} \#Loop eigenvalue storage
    \Function{Loop}{$v$ $\gets$ $\hat{v}$} \#For each previous result
        \State {$t = z_i$}
        \Function{Loop}{$m$ $\gets$ $\hat{m}$}
            \State{ $t \oplus (m_i)$ }
        \EndFunction
        \State{$\hat{u} = \neg \hat{m}$}    \# can ignore $Z_i$ not adjacent to $i$th qubit
        \State $\hat{c} = \sum_{n = 0}\binom{\hat{u}}{2n+t}$ \# combinations that give parity t
        \State{$\hat{m} = \hat{m} + Z_i$}
        \Function{Loop}{$c$ $\gets$ $\hat{c}$}
            \State{$\hat{e} = \hat{z} \lor c$}
            \State{$\hat{v_t} = \hat{v_t} + e$}
        \EndFunction
        \State{$\hat{v} = \hat{v_t}$}
    \EndFunction
\EndFunction

\State $\alpha_L =$ $\beta_L = 0$
\Function{Loop}{$v$ $\gets$ $\hat{v}$}
    \State{$j = bitwise\_sum(\hat{z})$}
    \State {$\hat{k} = \hat{v} \times \hat{x}$}
    \Function{Loop}{$k$ $\gets$ $\hat{k}$}
        \State $l = k \times B$ \# Parity check with logical operator
    	\State {\bf if} $l = 0$, $\quad \alpha_L$ = $\alpha_L + \alpha^j \beta^{N-j}$.
    	\State {\bf if} $l = 1$, $\quad \beta_L$ = $\beta_L + \alpha^j \beta^{N-j}$.
    \EndFunction
\EndFunction
\State \Return {$\{\alpha_L$, $\beta_L\}/\sqrt{|\alpha_L|^2+|\beta_L|^2}$}
\end{algorithmic}
\end{algorithm}

In this more performant algorithm, the goal is to determine which eigenvalues commute with the $Z$-sector of our trajectory and project these onto the target $X$-sector. The original algorithm projects our initial state through the $X$-stabilisers, which has an exponential number of terms in $d$, and only a fraction contribute to our encoded logical state for a specific $Z$-trajectory. By working backwards, we only need to reverse engineer the terms that will commute with our $Z$-stabiliser measurements. The $Z$-trajectory search is done in the loop from lines 3-19. For each $Z$-stabiliser, we can search for the combinations of the bits $= Z$ that XOR to give the correct value for our target. 

For example, consider a distance 2 unrotated surface code (Fig. \ref{fig:d2_unrot_layout}) with 5 data qubits and a target $\hat{0001}$. The two $Z$-stabilisers of the distance two surface code form the matrix below.
\begin{equation}
M = \begin{pmatrix}
D_1 & D_2 & D_3 & D_4 & D_5 \\
1 & 0 & 1 & 1 & 0 \\
0 & 1 & 1 & 0 & 1 
\end{pmatrix}
\label{eq:stabmatrix}
\end{equation}
If we create a representation of rows 3 and 4 (the $Z$-stabilisers) where we store the index of elements equal to one, we have a directory of which data qubits each $Z$-stabiliser impacts. In this example we use 0 as the first index.
\begin{equation}
M_{aux} = \begin{pmatrix}
0 & 2 & 3 \\
1 & 2 & 4
\end{pmatrix}
\end{equation}

For the first $Z$-stabiliser we look up the first bit of $\hat{01}$, which is $0$. Consider all combinations of $\{0, 2, 3\}$ that are of length $\{0, 2\}$ (or $\{1, 3\}$ if it were equal to $1$). We now iterate through each of $\{(\_), (0, 2), (0, 3), (2, 3)\}$ and set the bits of our initial trajectory $\{0\}_N$.

\begin{equation}
loop\_results = \begin{pmatrix}
0 & 0 & 0 & 0 & 0 \\
1 & 0 & 1 & 0 & 0 \\
1 & 0 & 0 & 1 & 0 \\
0 & 0 & 1 & 1 & 0 
\end{pmatrix}
\end{equation}

For each subsequent $Z$-stabiliser, we set $t$ equal to the $i$th bit of $\hat{01}$. Depending which values of $\{1, 2, 4\}$ have been set already, we create two subsets for the seen and unseen indices (i.e. $\{2\}$ and $\{1, 4\}$). For each result from the previous loop, we {\bf xor} $t$ with the value of each bit at the seen indices. As in the first case, we iterate through all combinations of unseen bits $\{1, 4\}$ that are odd or even lengths depending on $t$. Again we set the bits of the result at the indices — for each new combination, for each result from the previous loop. In our example below, $t$ will alternate between 0 and 1 as $z_2$ is our unseen and as $t$ is initialised as $1$ for the second bit, results will be flipped.

\begin{equation}
\begin{bmatrix}[c|ccccc|c] t & z_0 & z_1 & z_2 & z_3 & z_4 & $combinations$ \\
1 & 0 & 0 & 0 & 0 & 0 & (Z_{p_1}), (Z_{p_4})\\
0 & 1 & 0 & 1 & 0 & 0 & (\_), (Z_{p_1}, Z_{p_4})\\
1 & 1 & 0 & 0 & 1 & 0 & (Z_{p_1}), (Z_{p_4})\\
0 & 0 & 0 & 1 & 1 & 0 & (\_), (Z_{p_1}, Z_{p_4})
\end{bmatrix}
\end{equation}

\begin{equation}
Results = \begin{bmatrix}[ccccc|c] z_0 & z_1 & z_2 & z_3 & z_4 & $j$ \\
0 & 0 & 0 & 0 & 1 & 1 \\
0 & 1 & 0 & 0 & 0 & 1 \\
1 & 0 & 1 & 0 & 0 & 2 \\
1 & 1 & 1 & 0 & 1 & 4 \\
1 & 0 & 0 & 1 & 1 & 3 \\
1 & 1 & 0 & 1 & 0 & 3 \\
0 & 0 & 1 & 1 & 0 & 2 \\
0 & 1 & 1 & 1 & 1 & 4 
\end{bmatrix}
\end{equation}

For each row, we determine Hamming weight $j$ from the count of ones in $\hat{z}$. Note $j$ will remain the same during the next step even though the Hamming weight will change. 
\\
\\
\\
\\
It should be noted that after any loop in the algorithm, the intermediate results can be processed in parallel with no penalty in complexity aside from the negligible overhead of distributing the work. Using GPU acceleration, we have demonstrated sampling from the set of all trajectories with trivial $X$-syndromes for stabiliser codes with > 100 data qubits.
\\
\\
If we are now to examine a non-trivial $X$-syndrome, an additional step is needed. For example, lets examine a target trajectory of $\hat{1001}$ where the first two bits are our $X$-stabiliser measurement outcomes. We now project the results from our last step over these values as below.
\\

\begin{equation}
\begin{bmatrix}[c|cccc|c|c] \hat{z} & II & X_1 & X_2 & X_1 X_2 & $j$ & $k$\\
\hat{00001} & \hat{00001} & - \hat{11101} & \hat{00110} & - \hat{11010} & 1 & 0\\
\hat{01000} & \hat{01000} & - \hat{10100} & \hat{01111} & - \hat{10011} & 1 & 1\\
\hat{10100} & \hat{10100} & - \hat{01000} & \hat{10011} & - \hat{01111} & 2 & 1\\
\hat{11101} & \hat{11101} & - \hat{00001} & \hat{11010} & - \hat{00110} & 4 & 0\\
\hat{10011} & \hat{10011} & - \hat{01111} & \hat{10100} & - \hat{01000} & 3 & 1\\
\hat{11010} & \hat{11010} & - \hat{00110} & \hat{11101} & - \hat{00001} & 3 & 0\\
\hat{00110} & \hat{00110} & - \hat{11010} & \hat{00001} & - \hat{11101} & 2 & 0\\
\hat{01111} & \hat{01111} & - \hat{10011} & \hat{01000} & - \hat{10100} & 4 & 1
\end{bmatrix}
\end{equation}

We determine $k$ from the parity of the first $d$ bits (our logical observable, row 5 of Eq. \ref{eq:stabmatrix}) and these values can be used to construct the analytical equations (prior to re-normalisation) for our target. Each term contributes $\alpha^j \beta^{N-j}$ to the corresponding logical state, and we can collect them as below.
\\
\\
\begin{equation}
\begin{aligned}
\begin{pmatrix} \ket{00001} \\ \ket{00110} \\ \ket{11101} \\ \ket{11010}  \\ \ket{01000}  \\ \ket{01111}  \\ \ket{10100}  \\ \ket{10011}\end{pmatrix} = \begin{pmatrix} 
2\alpha^4\beta + 2\alpha^3\beta^2  - 2\alpha^2\beta^3 - 2\alpha\beta^4 \\ 
2\alpha^4\beta + 2\alpha^3\beta^2  - 2\alpha^2\beta^3 - 2\alpha\beta^4 \\ 
-2\alpha^4\beta - 2\alpha^3\beta^2  + 2\alpha^2\beta^3 + 2\alpha\beta^4 \\ 
-2\alpha^4\beta - 2\alpha^3\beta^2  + 2\alpha^2\beta^3 + 2\alpha\beta^4 \\ 
2\alpha^4\beta - 2\alpha^3\beta^2  - 2\alpha^2\beta^3 + 2\alpha\beta^4 \\ 
2\alpha^4\beta - 2\alpha^3\beta^2  - 2\alpha^2\beta^3 + 2\alpha\beta^4 \\ 
- 2\alpha^4\beta + 2\alpha^3\beta^2  + 2\alpha^2\beta^3 - 2\alpha\beta^4 \\ 
- 2\alpha^4\beta + 2\alpha^3\beta^2  + 2\alpha^2\beta^3 - 2\alpha\beta^4 \end{pmatrix} \\
\end{aligned}
\label{eq:4eqn}
\end{equation}
\\
\\

Due to the Pauli frame set by our stabiliser measurements, what we consider our logical state is a mix of the following terms.

\begin{equation}
\begin{aligned}
\ket{\Lambda} = \begin{pmatrix} \alpha_L \\ \beta_L \end{pmatrix} = \begin{pmatrix} 
\ket{00001} + \ket{00110} - \ket{11101} - \ket{11010}  \\ - \ket{01000}  - \ket{01111}  + \ket{10100}  + \ket{10011} \end{pmatrix} \\
\\
\ket{\Lambda} = \begin{pmatrix} 
\alpha^4\beta + \alpha^3\beta^2  - \alpha^2\beta^3 - \alpha\beta^4 \\  - \alpha^4\beta + \alpha^3\beta^2  + \alpha^2\beta^3 - \alpha\beta^4 \end{pmatrix} \\
\end{aligned}
\label{eq:5eqn}
\end{equation}
\\
\\
These additional steps for calculating for non-trivial $X$-stabiliser measurements are costly and this classical algorithm is only feasible for code distances < 6.

\section{Appendix B - Functional forms for $d=3$ code}
For $N=13$ data qubits, there are a total of $2^{(13-1)/2} = 64$ $Z$-stabiliser trajectories.  Each of these trajectories is specified by the 6-bit, $Z$ parity vector, and each stabiliser trajectory results in a different logical state as a function of $|\chi\rangle^
{\otimes N} = \left(\alpha|0\rangle+\beta|1\rangle\right)^{\otimes N}$. These functional forms are {\em not} normalised.
\\
\\
When initialising a $d=3$ planar surface code, the specific Pauli frame of the $Z$-stabilisers that are determined after decoding allows you to calculate the actual logically encoded state.
\begin{equation}
\begin{aligned}
\hat{000000} = &\begin{pmatrix} \alpha^{13} + 4\alpha^{10}\beta^{3} + 4\alpha^{9}\beta^{4} + 4\alpha^{8}\beta^{5} + 14\alpha^{7}\beta^{6} + 20\alpha^{6}\beta^{7} + 11\alpha^{5}\beta^{8} + 4\alpha^{4}\beta^{9} + 2\alpha^{3}\beta^{10} \\ 3\alpha^{10}\beta^{3} + 8\alpha^{9}\beta^{4} + 10\alpha^{8}\beta^{5} + 8\alpha^{7}\beta^{6} + 12\alpha^{6}\beta^{7} + 16\alpha^{5}\beta^{8} + 6\alpha^{4}\beta^{9} + \alpha^{2}\beta^{11}\end{pmatrix} \\
\hat{000001} = &\begin{pmatrix} \alpha^{12}\beta + \alpha^{11}\beta^{2} + \alpha^{10}\beta^{3} + 4\alpha^{9}\beta^{4} + 10\alpha^{8}\beta^{5} + 14\alpha^{7}\beta^{6} + 14\alpha^{6}\beta^{7} + 12\alpha^{5}\beta^{8} + 5\alpha^{4}\beta^{9} + \alpha^{3}\beta^{10} + \alpha^{2}\beta^{11} \\ \alpha^{11}\beta^{2} + 2\alpha^{10}\beta^{3} + 6\alpha^{9}\beta^{4} + 10\alpha^{8}\beta^{5} + 12\alpha^{7}\beta^{6} + 14\alpha^{6}\beta^{7} + 10\alpha^{5}\beta^{8} + 6\alpha^{4}\beta^{9} + 3\alpha^{3}\beta^{10}\end{pmatrix} \\
\hat{000010} = &\begin{pmatrix} \alpha^{12}\beta + 2\alpha^{11}\beta^{2} + 4\alpha^{9}\beta^{4} + 10\alpha^{8}\beta^{5} + 12\alpha^{7}\beta^{6} + 16\alpha^{6}\beta^{7} + 12\alpha^{5}\beta^{8} + 5\alpha^{4}\beta^{9} + 2\alpha^{3}\beta^{10} \\ \alpha^{11}\beta^{2} + 4\alpha^{10}\beta^{3} + 4\alpha^{9}\beta^{4} + 8\alpha^{8}\beta^{5} + 14\alpha^{7}\beta^{6} + 12\alpha^{6}\beta^{7} + 12\alpha^{5}\beta^{8} + 8\alpha^{4}\beta^{9} + \alpha^{3}\beta^{10}\end{pmatrix} \\
\hat{000011} = &\begin{pmatrix} \alpha^{12}\beta + \alpha^{11}\beta^{2} + 2\alpha^{10}\beta^{3} + 4\alpha^{9}\beta^{4} + 8\alpha^{8}\beta^{5} + 14\alpha^{7}\beta^{6} + 14\alpha^{6}\beta^{7} + 12\alpha^{5}\beta^{8} + 7\alpha^{4}\beta^{9} + \alpha^{3}\beta^{10} \\ 2\alpha^{10}\beta^{3} + 8\alpha^{9}\beta^{4} + 10\alpha^{8}\beta^{5} + 12\alpha^{7}\beta^{6} + 14\alpha^{6}\beta^{7} + 8\alpha^{5}\beta^{8} + 6\alpha^{4}\beta^{9} + 4\alpha^{3}\beta^{10}\end{pmatrix} \\
\hat{000100} = &\begin{pmatrix} \alpha^{12}\beta + \alpha^{11}\beta^{2} + \alpha^{10}\beta^{3} + 4\alpha^{9}\beta^{4} + 10\alpha^{8}\beta^{5} + 14\alpha^{7}\beta^{6} + 14\alpha^{6}\beta^{7} + 12\alpha^{5}\beta^{8} + 5\alpha^{4}\beta^{9} + \alpha^{3}\beta^{10} + \alpha^{2}\beta^{11} \\ \alpha^{11}\beta^{2} + 2\alpha^{10}\beta^{3} + 6\alpha^{9}\beta^{4} + 10\alpha^{8}\beta^{5} + 12\alpha^{7}\beta^{6} + 14\alpha^{6}\beta^{7} + 10\alpha^{5}\beta^{8} + 6\alpha^{4}\beta^{9} + 3\alpha^{3}\beta^{10}\end{pmatrix} \\
\hat{000101} = &\begin{pmatrix} 2\alpha^{11}\beta^{2} + 2\alpha^{10}\beta^{3} + 4\alpha^{9}\beta^{4} + 10\alpha^{8}\beta^{5} + 12\alpha^{7}\beta^{6} + 14\alpha^{6}\beta^{7} + 12\alpha^{5}\beta^{8} + 6\alpha^{4}\beta^{9} + 2\alpha^{3}\beta^{10} \\ 2\alpha^{10}\beta^{3} + 6\alpha^{9}\beta^{4} + 12\alpha^{8}\beta^{5} + 14\alpha^{7}\beta^{6} + 12\alpha^{6}\beta^{7} + 10\alpha^{5}\beta^{8} + 4\alpha^{4}\beta^{9} + 2\alpha^{3}\beta^{10} + 2\alpha^{2}\beta^{11}\end{pmatrix} \\
\hat{000110} = &\begin{pmatrix} \alpha^{12}\beta + \alpha^{11}\beta^{2} + 2\alpha^{10}\beta^{3} + 4\alpha^{9}\beta^{4} + 8\alpha^{8}\beta^{5} + 14\alpha^{7}\beta^{6} + 14\alpha^{6}\beta^{7} + 12\alpha^{5}\beta^{8} + 7\alpha^{4}\beta^{9} + \alpha^{3}\beta^{10} \\ 2\alpha^{10}\beta^{3} + 8\alpha^{9}\beta^{4} + 10\alpha^{8}\beta^{5} + 12\alpha^{7}\beta^{6} + 14\alpha^{6}\beta^{7} + 8\alpha^{5}\beta^{8} + 6\alpha^{4}\beta^{9} + 4\alpha^{3}\beta^{10}\end{pmatrix} \\
\hat{000111} = &\begin{pmatrix} 2\alpha^{11}\beta^{2} + 2\alpha^{10}\beta^{3} + 2\alpha^{9}\beta^{4} + 10\alpha^{8}\beta^{5} + 18\alpha^{7}\beta^{6} + 14\alpha^{6}\beta^{7} + 6\alpha^{5}\beta^{8} + 6\alpha^{4}\beta^{9} + 4\alpha^{3}\beta^{10} \\ 2\alpha^{10}\beta^{3} + 6\alpha^{9}\beta^{4} + 14\alpha^{8}\beta^{5} + 14\alpha^{7}\beta^{6} + 6\alpha^{6}\beta^{7} + 10\alpha^{5}\beta^{8} + 10\alpha^{4}\beta^{9} + 2\alpha^{3}\beta^{10}\end{pmatrix} \\
\hat{001000} = &\begin{pmatrix} \alpha^{11}\beta^{2} + 2\alpha^{10}\beta^{3} + 6\alpha^{9}\beta^{4} + 10\alpha^{8}\beta^{5} + 12\alpha^{7}\beta^{6} + 14\alpha^{6}\beta^{7} + 10\alpha^{5}\beta^{8} + 6\alpha^{4}\beta^{9} + 3\alpha^{3}\beta^{10} \\ \alpha^{12}\beta + \alpha^{11}\beta^{2} + \alpha^{10}\beta^{3} + 4\alpha^{9}\beta^{4} + 10\alpha^{8}\beta^{5} + 14\alpha^{7}\beta^{6} + 14\alpha^{6}\beta^{7} + 12\alpha^{5}\beta^{8} + 5\alpha^{4}\beta^{9} + \alpha^{3}\beta^{10} + \alpha^{2}\beta^{11}\end{pmatrix} \\
\hat{001001} = &\begin{pmatrix} \alpha^{12}\beta + \alpha^{10}\beta^{3} + 6\alpha^{9}\beta^{4} + 10\alpha^{8}\beta^{5} + 14\alpha^{7}\beta^{6} + 14\alpha^{6}\beta^{7} + 10\alpha^{5}\beta^{8} + 5\alpha^{4}\beta^{9} + 2\alpha^{3}\beta^{10} + \alpha^{2}\beta^{11} \\ \alpha^{11}\beta^{2} + 2\alpha^{10}\beta^{3} + 5\alpha^{9}\beta^{4} + 10\alpha^{8}\beta^{5} + 14\alpha^{7}\beta^{6} + 14\alpha^{6}\beta^{7} + 10\alpha^{5}\beta^{8} + 6\alpha^{4}\beta^{9} + \alpha^{3}\beta^{10} + \alpha\beta^{12}\end{pmatrix} \\
\hat{001010} = &\begin{pmatrix} 2\alpha^{11}\beta^{2} + 2\alpha^{10}\beta^{3} + 4\alpha^{9}\beta^{4} + 10\alpha^{8}\beta^{5} + 12\alpha^{7}\beta^{6} + 14\alpha^{6}\beta^{7} + 12\alpha^{5}\beta^{8} + 6\alpha^{4}\beta^{9} + 2\alpha^{3}\beta^{10} \\ \alpha^{11}\beta^{2} + 3\alpha^{10}\beta^{3} + 4\alpha^{9}\beta^{4} + 10\alpha^{8}\beta^{5} + 14\alpha^{7}\beta^{6} + 12\alpha^{6}\beta^{7} + 12\alpha^{5}\beta^{8} + 6\alpha^{4}\beta^{9} + \alpha^{3}\beta^{10} + \alpha^{2}\beta^{11}\end{pmatrix} \\
\hat{001011} = &\begin{pmatrix} \alpha^{11}\beta^{2} + 3\alpha^{10}\beta^{3} + 2\alpha^{9}\beta^{4} + 10\alpha^{8}\beta^{5} + 20\alpha^{7}\beta^{6} + 12\alpha^{6}\beta^{7} + 6\alpha^{5}\beta^{8} + 6\alpha^{4}\beta^{9} + 3\alpha^{3}\beta^{10} + \alpha^{2}\beta^{11} \\ \alpha^{11}\beta^{2} + 3\alpha^{10}\beta^{3} + 6\alpha^{9}\beta^{4} + 6\alpha^{8}\beta^{5} + 12\alpha^{7}\beta^{6} + 20\alpha^{6}\beta^{7} + 10\alpha^{5}\beta^{8} + 2\alpha^{4}\beta^{9} + 3\alpha^{3}\beta^{10} + \alpha^{2}\beta^{11}\end{pmatrix} \\
\hat{001100} = &\begin{pmatrix} 5\alpha^{10}\beta^{3} + 6\alpha^{9}\beta^{4} + 4\alpha^{8}\beta^{5} + 14\alpha^{7}\beta^{6} + 18\alpha^{6}\beta^{7} + 10\alpha^{5}\beta^{8} + 4\alpha^{4}\beta^{9} + 2\alpha^{3}\beta^{10} + \alpha^{2}\beta^{11} \\ \alpha^{11}\beta^{2} + 2\alpha^{10}\beta^{3} + 4\alpha^{9}\beta^{4} + 10\alpha^{8}\beta^{5} + 18\alpha^{7}\beta^{6} + 14\alpha^{6}\beta^{7} + 4\alpha^{5}\beta^{8} + 6\alpha^{4}\beta^{9} + 5\alpha^{3}\beta^{10}\end{pmatrix} \\
\hat{001101} = &\begin{pmatrix} \alpha^{11}\beta^{2} + \alpha^{10}\beta^{3} + 5\alpha^{9}\beta^{4} + 12\alpha^{8}\beta^{5} + 14\alpha^{7}\beta^{6} + 14\alpha^{6}\beta^{7} + 10\alpha^{5}\beta^{8} + 4\alpha^{4}\beta^{9} + \alpha^{3}\beta^{10} + \alpha^{2}\beta^{11} + \alpha\beta^{12} \\ 3\alpha^{10}\beta^{3} + 6\alpha^{9}\beta^{4} + 10\alpha^{8}\beta^{5} + 14\alpha^{7}\beta^{6} + 12\alpha^{6}\beta^{7} + 10\alpha^{5}\beta^{8} + 6\alpha^{4}\beta^{9} + 2\alpha^{3}\beta^{10} + \alpha^{2}\beta^{11}\end{pmatrix} \\
\hat{001110} = &\begin{pmatrix} 3\alpha^{10}\beta^{3} + 6\alpha^{9}\beta^{4} + 10\alpha^{8}\beta^{5} + 14\alpha^{7}\beta^{6} + 12\alpha^{6}\beta^{7} + 10\alpha^{5}\beta^{8} + 6\alpha^{4}\beta^{9} + 2\alpha^{3}\beta^{10} + \alpha^{2}\beta^{11} \\ \alpha^{11}\beta^{2} + 2\alpha^{10}\beta^{3} + 6\alpha^{9}\beta^{4} + 10\alpha^{8}\beta^{5} + 12\alpha^{7}\beta^{6} + 14\alpha^{6}\beta^{7} + 10\alpha^{5}\beta^{8} + 6\alpha^{4}\beta^{9} + 3\alpha^{3}\beta^{10}\end{pmatrix} \\
\hat{001111} = &\begin{pmatrix} \alpha^{11}\beta^{2} + \alpha^{10}\beta^{3} + 6\alpha^{9}\beta^{4} + 12\alpha^{8}\beta^{5} + 12\alpha^{7}\beta^{6} + 14\alpha^{6}\beta^{7} + 10\alpha^{5}\beta^{8} + 4\alpha^{4}\beta^{9} + 3\alpha^{3}\beta^{10} + \alpha^{2}\beta^{11} \\ 2\alpha^{10}\beta^{3} + 6\alpha^{9}\beta^{4} + 12\alpha^{8}\beta^{5} + 14\alpha^{7}\beta^{6} + 12\alpha^{6}\beta^{7} + 10\alpha^{5}\beta^{8} + 4\alpha^{4}\beta^{9} + 2\alpha^{3}\beta^{10} + 2\alpha^{2}\beta^{11}\end{pmatrix} \\
\hat{010000} = &\begin{pmatrix} \alpha^{11}\beta^{2} + 4\alpha^{10}\beta^{3} + 4\alpha^{9}\beta^{4} + 8\alpha^{8}\beta^{5} + 14\alpha^{7}\beta^{6} + 12\alpha^{6}\beta^{7} + 12\alpha^{5}\beta^{8} + 8\alpha^{4}\beta^{9} + \alpha^{3}\beta^{10} \\ \alpha^{12}\beta + 2\alpha^{11}\beta^{2} + 4\alpha^{9}\beta^{4} + 10\alpha^{8}\beta^{5} + 12\alpha^{7}\beta^{6} + 16\alpha^{6}\beta^{7} + 12\alpha^{5}\beta^{8} + 5\alpha^{4}\beta^{9} + 2\alpha^{3}\beta^{10}\end{pmatrix} \\
\hat{010001} = &\begin{pmatrix} 2\alpha^{11}\beta^{2} + 2\alpha^{10}\beta^{3} + 4\alpha^{9}\beta^{4} + 10\alpha^{8}\beta^{5} + 12\alpha^{7}\beta^{6} + 14\alpha^{6}\beta^{7} + 12\alpha^{5}\beta^{8} + 6\alpha^{4}\beta^{9} + 2\alpha^{3}\beta^{10} \\ \alpha^{11}\beta^{2} + 3\alpha^{10}\beta^{3} + 4\alpha^{9}\beta^{4} + 10\alpha^{8}\beta^{5} + 14\alpha^{7}\beta^{6} + 12\alpha^{6}\beta^{7} + 12\alpha^{5}\beta^{8} + 6\alpha^{4}\beta^{9} + \alpha^{3}\beta^{10} + \alpha^{2}\beta^{11}\end{pmatrix} \\
\hat{010010} = &\begin{pmatrix} \alpha^{12}\beta + 2\alpha^{10}\beta^{3} + 8\alpha^{9}\beta^{4} + 8\alpha^{8}\beta^{5} + 8\alpha^{7}\beta^{6} + 14\alpha^{6}\beta^{7} + 16\alpha^{5}\beta^{8} + 7\alpha^{4}\beta^{9} \\ \alpha^{11}\beta^{2} + 4\alpha^{10}\beta^{3} + 6\alpha^{9}\beta^{4} + 4\alpha^{8}\beta^{5} + 12\alpha^{7}\beta^{6} + 20\alpha^{6}\beta^{7} + 10\alpha^{5}\beta^{8} + 4\alpha^{4}\beta^{9} + 3\alpha^{3}\beta^{10}\end{pmatrix} \\
\hat{010011} = &\begin{pmatrix} \alpha^{11}\beta^{2} + 2\alpha^{10}\beta^{3} + 6\alpha^{9}\beta^{4} + 10\alpha^{8}\beta^{5} + 12\alpha^{7}\beta^{6} + 14\alpha^{6}\beta^{7} + 10\alpha^{5}\beta^{8} + 6\alpha^{4}\beta^{9} + 3\alpha^{3}\beta^{10} \\ \alpha^{11}\beta^{2} + 4\alpha^{10}\beta^{3} + 4\alpha^{9}\beta^{4} + 8\alpha^{8}\beta^{5} + 14\alpha^{7}\beta^{6} + 12\alpha^{6}\beta^{7} + 12\alpha^{5}\beta^{8} + 8\alpha^{4}\beta^{9} + \alpha^{3}\beta^{10}\end{pmatrix} \\
\hat{010100} = &\begin{pmatrix} 2\alpha^{11}\beta^{2} + 2\alpha^{10}\beta^{3} + 4\alpha^{9}\beta^{4} + 10\alpha^{8}\beta^{5} + 12\alpha^{7}\beta^{6} + 14\alpha^{6}\beta^{7} + 12\alpha^{5}\beta^{8} + 6\alpha^{4}\beta^{9} + 2\alpha^{3}\beta^{10} \\ \alpha^{11}\beta^{2} + 3\alpha^{10}\beta^{3} + 4\alpha^{9}\beta^{4} + 10\alpha^{8}\beta^{5} + 14\alpha^{7}\beta^{6} + 12\alpha^{6}\beta^{7} + 12\alpha^{5}\beta^{8} + 6\alpha^{4}\beta^{9} + \alpha^{3}\beta^{10} + \alpha^{2}\beta^{11}\end{pmatrix} \\
\hat{010101} = &\begin{pmatrix} 4\alpha^{10}\beta^{3} + 6\alpha^{9}\beta^{4} + 6\alpha^{8}\beta^{5} + 14\alpha^{7}\beta^{6} + 18\alpha^{6}\beta^{7} + 10\alpha^{5}\beta^{8} + 2\alpha^{4}\beta^{9} + 2\alpha^{3}\beta^{10} + 2\alpha^{2}\beta^{11} \\ 2\alpha^{10}\beta^{3} + 10\alpha^{9}\beta^{4} + 10\alpha^{8}\beta^{5} + 6\alpha^{7}\beta^{6} + 14\alpha^{6}\beta^{7} + 14\alpha^{5}\beta^{8} + 6\alpha^{4}\beta^{9} + 2\alpha^{3}\beta^{10}\end{pmatrix} \\
\end{aligned}
\end{equation}
\begin{equation}
\begin{aligned}
\hat{010110} = &\begin{pmatrix} \alpha^{11}\beta^{2} + 2\alpha^{10}\beta^{3} + 6\alpha^{9}\beta^{4} + 10\alpha^{8}\beta^{5} + 12\alpha^{7}\beta^{6} + 14\alpha^{6}\beta^{7} + 10\alpha^{5}\beta^{8} + 6\alpha^{4}\beta^{9} + 3\alpha^{3}\beta^{10} \\ \alpha^{11}\beta^{2} + 4\alpha^{10}\beta^{3} + 4\alpha^{9}\beta^{4} + 8\alpha^{8}\beta^{5} + 14\alpha^{7}\beta^{6} + 12\alpha^{6}\beta^{7} + 12\alpha^{5}\beta^{8} + 8\alpha^{4}\beta^{9} + \alpha^{3}\beta^{10}\end{pmatrix} \\
\hat{010111} = &\begin{pmatrix} 4\alpha^{10}\beta^{3} + 6\alpha^{9}\beta^{4} + 8\alpha^{8}\beta^{5} + 14\alpha^{7}\beta^{6} + 12\alpha^{6}\beta^{7} + 10\alpha^{5}\beta^{8} + 8\alpha^{4}\beta^{9} + 2\alpha^{3}\beta^{10} \\ 2\alpha^{10}\beta^{3} + 8\alpha^{9}\beta^{4} + 10\alpha^{8}\beta^{5} + 12\alpha^{7}\beta^{6} + 14\alpha^{6}\beta^{7} + 8\alpha^{5}\beta^{8} + 6\alpha^{4}\beta^{9} + 4\alpha^{3}\beta^{10}\end{pmatrix} \\
\hat{011000} = &\begin{pmatrix} \alpha^{12}\beta + \alpha^{11}\beta^{2} + 2\alpha^{10}\beta^{3} + 4\alpha^{9}\beta^{4} + 8\alpha^{8}\beta^{5} + 14\alpha^{7}\beta^{6} + 14\alpha^{6}\beta^{7} + 12\alpha^{5}\beta^{8} + 7\alpha^{4}\beta^{9} + \alpha^{3}\beta^{10} \\ 2\alpha^{10}\beta^{3} + 8\alpha^{9}\beta^{4} + 10\alpha^{8}\beta^{5} + 12\alpha^{7}\beta^{6} + 14\alpha^{6}\beta^{7} + 8\alpha^{5}\beta^{8} + 6\alpha^{4}\beta^{9} + 4\alpha^{3}\beta^{10}\end{pmatrix} \\
\hat{011001} = &\begin{pmatrix} \alpha^{11}\beta^{2} + 3\alpha^{10}\beta^{3} + 6\alpha^{9}\beta^{4} + 6\alpha^{8}\beta^{5} + 12\alpha^{7}\beta^{6} + 20\alpha^{6}\beta^{7} + 10\alpha^{5}\beta^{8} + 2\alpha^{4}\beta^{9} + 3\alpha^{3}\beta^{10} + \alpha^{2}\beta^{11} \\ \alpha^{11}\beta^{2} + 3\alpha^{10}\beta^{3} + 2\alpha^{9}\beta^{4} + 10\alpha^{8}\beta^{5} + 20\alpha^{7}\beta^{6} + 12\alpha^{6}\beta^{7} + 6\alpha^{5}\beta^{8} + 6\alpha^{4}\beta^{9} + 3\alpha^{3}\beta^{10} + \alpha^{2}\beta^{11}\end{pmatrix} \\
\hat{011010} = &\begin{pmatrix} \alpha^{11}\beta^{2} + 4\alpha^{10}\beta^{3} + 4\alpha^{9}\beta^{4} + 8\alpha^{8}\beta^{5} + 14\alpha^{7}\beta^{6} + 12\alpha^{6}\beta^{7} + 12\alpha^{5}\beta^{8} + 8\alpha^{4}\beta^{9} + \alpha^{3}\beta^{10} \\ \alpha^{11}\beta^{2} + 2\alpha^{10}\beta^{3} + 6\alpha^{9}\beta^{4} + 10\alpha^{8}\beta^{5} + 12\alpha^{7}\beta^{6} + 14\alpha^{6}\beta^{7} + 10\alpha^{5}\beta^{8} + 6\alpha^{4}\beta^{9} + 3\alpha^{3}\beta^{10}\end{pmatrix} \\
\hat{011011} = &\begin{pmatrix} 2\alpha^{11}\beta^{2} + 2\alpha^{10}\beta^{3} + 4\alpha^{9}\beta^{4} + 10\alpha^{8}\beta^{5} + 12\alpha^{7}\beta^{6} + 14\alpha^{6}\beta^{7} + 12\alpha^{5}\beta^{8} + 6\alpha^{4}\beta^{9} + 2\alpha^{3}\beta^{10} \\ 2\alpha^{10}\beta^{3} + 6\alpha^{9}\beta^{4} + 12\alpha^{8}\beta^{5} + 14\alpha^{7}\beta^{6} + 12\alpha^{6}\beta^{7} + 10\alpha^{5}\beta^{8} + 4\alpha^{4}\beta^{9} + 2\alpha^{3}\beta^{10} + 2\alpha^{2}\beta^{11}\end{pmatrix} \\
\hat{011100} = &\begin{pmatrix} \alpha^{11}\beta^{2} + 2\alpha^{10}\beta^{3} + 6\alpha^{9}\beta^{4} + 10\alpha^{8}\beta^{5} + 12\alpha^{7}\beta^{6} + 14\alpha^{6}\beta^{7} + 10\alpha^{5}\beta^{8} + 6\alpha^{4}\beta^{9} + 3\alpha^{3}\beta^{10} \\ 3\alpha^{10}\beta^{3} + 6\alpha^{9}\beta^{4} + 10\alpha^{8}\beta^{5} + 14\alpha^{7}\beta^{6} + 12\alpha^{6}\beta^{7} + 10\alpha^{5}\beta^{8} + 6\alpha^{4}\beta^{9} + 2\alpha^{3}\beta^{10} + \alpha^{2}\beta^{11}\end{pmatrix} \\
\hat{011101} = &\begin{pmatrix} 4\alpha^{10}\beta^{3} + 6\alpha^{9}\beta^{4} + 8\alpha^{8}\beta^{5} + 14\alpha^{7}\beta^{6} + 12\alpha^{6}\beta^{7} + 10\alpha^{5}\beta^{8} + 8\alpha^{4}\beta^{9} + 2\alpha^{3}\beta^{10} \\ \alpha^{10}\beta^{3} + 7\alpha^{9}\beta^{4} + 12\alpha^{8}\beta^{5} + 14\alpha^{7}\beta^{6} + 14\alpha^{6}\beta^{7} + 8\alpha^{5}\beta^{8} + 4\alpha^{4}\beta^{9} + 2\alpha^{3}\beta^{10} + \alpha^{2}\beta^{11} + \alpha\beta^{12}\end{pmatrix} \\
\hat{011110} = &\begin{pmatrix} \alpha^{11}\beta^{2} + 2\alpha^{10}\beta^{3} + 8\alpha^{9}\beta^{4} + 10\alpha^{8}\beta^{5} + 6\alpha^{7}\beta^{6} + 14\alpha^{6}\beta^{7} + 16\alpha^{5}\beta^{8} + 6\alpha^{4}\beta^{9} + \alpha^{3}\beta^{10} \\ \alpha^{10}\beta^{3} + 6\alpha^{9}\beta^{4} + 16\alpha^{8}\beta^{5} + 14\alpha^{7}\beta^{6} + 6\alpha^{6}\beta^{7} + 10\alpha^{5}\beta^{8} + 8\alpha^{4}\beta^{9} + 2\alpha^{3}\beta^{10} + \alpha^{2}\beta^{11}\end{pmatrix} \\
\hat{011111} = &\begin{pmatrix} 3\alpha^{10}\beta^{3} + 6\alpha^{9}\beta^{4} + 10\alpha^{8}\beta^{5} + 14\alpha^{7}\beta^{6} + 12\alpha^{6}\beta^{7} + 10\alpha^{5}\beta^{8} + 6\alpha^{4}\beta^{9} + 2\alpha^{3}\beta^{10} + \alpha^{2}\beta^{11} \\ \alpha^{10}\beta^{3} + 8\alpha^{9}\beta^{4} + 12\alpha^{8}\beta^{5} + 12\alpha^{7}\beta^{6} + 14\alpha^{6}\beta^{7} + 8\alpha^{5}\beta^{8} + 4\alpha^{4}\beta^{9} + 4\alpha^{3}\beta^{10} + \alpha^{2}\beta^{11}\end{pmatrix} \\
\hat{100000} = &\begin{pmatrix} \alpha^{11}\beta^{2} + 2\alpha^{10}\beta^{3} + 6\alpha^{9}\beta^{4} + 10\alpha^{8}\beta^{5} + 12\alpha^{7}\beta^{6} + 14\alpha^{6}\beta^{7} + 10\alpha^{5}\beta^{8} + 6\alpha^{4}\beta^{9} + 3\alpha^{3}\beta^{10} \\ \alpha^{12}\beta + \alpha^{11}\beta^{2} + \alpha^{10}\beta^{3} + 4\alpha^{9}\beta^{4} + 10\alpha^{8}\beta^{5} + 14\alpha^{7}\beta^{6} + 14\alpha^{6}\beta^{7} + 12\alpha^{5}\beta^{8} + 5\alpha^{4}\beta^{9} + \alpha^{3}\beta^{10} + \alpha^{2}\beta^{11}\end{pmatrix} \\
\hat{100001} = &\begin{pmatrix} 5\alpha^{10}\beta^{3} + 6\alpha^{9}\beta^{4} + 4\alpha^{8}\beta^{5} + 14\alpha^{7}\beta^{6} + 18\alpha^{6}\beta^{7} + 10\alpha^{5}\beta^{8} + 4\alpha^{4}\beta^{9} + 2\alpha^{3}\beta^{10} + \alpha^{2}\beta^{11} \\ \alpha^{11}\beta^{2} + 2\alpha^{10}\beta^{3} + 4\alpha^{9}\beta^{4} + 10\alpha^{8}\beta^{5} + 18\alpha^{7}\beta^{6} + 14\alpha^{6}\beta^{7} + 4\alpha^{5}\beta^{8} + 6\alpha^{4}\beta^{9} + 5\alpha^{3}\beta^{10}\end{pmatrix} \\
\hat{100010} = &\begin{pmatrix} 2\alpha^{11}\beta^{2} + 2\alpha^{10}\beta^{3} + 4\alpha^{9}\beta^{4} + 10\alpha^{8}\beta^{5} + 12\alpha^{7}\beta^{6} + 14\alpha^{6}\beta^{7} + 12\alpha^{5}\beta^{8} + 6\alpha^{4}\beta^{9} + 2\alpha^{3}\beta^{10} \\ \alpha^{11}\beta^{2} + 3\alpha^{10}\beta^{3} + 4\alpha^{9}\beta^{4} + 10\alpha^{8}\beta^{5} + 14\alpha^{7}\beta^{6} + 12\alpha^{6}\beta^{7} + 12\alpha^{5}\beta^{8} + 6\alpha^{4}\beta^{9} + \alpha^{3}\beta^{10} + \alpha^{2}\beta^{11}\end{pmatrix} \\
\hat{100011} = &\begin{pmatrix} 3\alpha^{10}\beta^{3} + 6\alpha^{9}\beta^{4} + 10\alpha^{8}\beta^{5} + 14\alpha^{7}\beta^{6} + 12\alpha^{6}\beta^{7} + 10\alpha^{5}\beta^{8} + 6\alpha^{4}\beta^{9} + 2\alpha^{3}\beta^{10} + \alpha^{2}\beta^{11} \\ \alpha^{11}\beta^{2} + 2\alpha^{10}\beta^{3} + 6\alpha^{9}\beta^{4} + 10\alpha^{8}\beta^{5} + 12\alpha^{7}\beta^{6} + 14\alpha^{6}\beta^{7} + 10\alpha^{5}\beta^{8} + 6\alpha^{4}\beta^{9} + 3\alpha^{3}\beta^{10}\end{pmatrix} \\
\hat{100100} = &\begin{pmatrix} \alpha^{12}\beta + \alpha^{10}\beta^{3} + 6\alpha^{9}\beta^{4} + 10\alpha^{8}\beta^{5} + 14\alpha^{7}\beta^{6} + 14\alpha^{6}\beta^{7} + 10\alpha^{5}\beta^{8} + 5\alpha^{4}\beta^{9} + 2\alpha^{3}\beta^{10} + \alpha^{2}\beta^{11} \\ \alpha^{11}\beta^{2} + 2\alpha^{10}\beta^{3} + 5\alpha^{9}\beta^{4} + 10\alpha^{8}\beta^{5} + 14\alpha^{7}\beta^{6} + 14\alpha^{6}\beta^{7} + 10\alpha^{5}\beta^{8} + 6\alpha^{4}\beta^{9} + \alpha^{3}\beta^{10} + \alpha\beta^{12}\end{pmatrix} \\
\hat{100101} = &\begin{pmatrix} \alpha^{11}\beta^{2} + \alpha^{10}\beta^{3} + 5\alpha^{9}\beta^{4} + 12\alpha^{8}\beta^{5} + 14\alpha^{7}\beta^{6} + 14\alpha^{6}\beta^{7} + 10\alpha^{5}\beta^{8} + 4\alpha^{4}\beta^{9} + \alpha^{3}\beta^{10} + \alpha^{2}\beta^{11} + \alpha\beta^{12} \\ 3\alpha^{10}\beta^{3} + 6\alpha^{9}\beta^{4} + 10\alpha^{8}\beta^{5} + 14\alpha^{7}\beta^{6} + 12\alpha^{6}\beta^{7} + 10\alpha^{5}\beta^{8} + 6\alpha^{4}\beta^{9} + 2\alpha^{3}\beta^{10} + \alpha^{2}\beta^{11}\end{pmatrix} \\
\hat{100110} = &\begin{pmatrix} \alpha^{11}\beta^{2} + 3\alpha^{10}\beta^{3} + 2\alpha^{9}\beta^{4} + 10\alpha^{8}\beta^{5} + 20\alpha^{7}\beta^{6} + 12\alpha^{6}\beta^{7} + 6\alpha^{5}\beta^{8} + 6\alpha^{4}\beta^{9} + 3\alpha^{3}\beta^{10} + \alpha^{2}\beta^{11} \\ \alpha^{11}\beta^{2} + 3\alpha^{10}\beta^{3} + 6\alpha^{9}\beta^{4} + 6\alpha^{8}\beta^{5} + 12\alpha^{7}\beta^{6} + 20\alpha^{6}\beta^{7} + 10\alpha^{5}\beta^{8} + 2\alpha^{4}\beta^{9} + 3\alpha^{3}\beta^{10} + \alpha^{2}\beta^{11}\end{pmatrix} \\
\hat{100111} = &\begin{pmatrix} \alpha^{11}\beta^{2} + \alpha^{10}\beta^{3} + 6\alpha^{9}\beta^{4} + 12\alpha^{8}\beta^{5} + 12\alpha^{7}\beta^{6} + 14\alpha^{6}\beta^{7} + 10\alpha^{5}\beta^{8} + 4\alpha^{4}\beta^{9} + 3\alpha^{3}\beta^{10} + \alpha^{2}\beta^{11} \\ 2\alpha^{10}\beta^{3} + 6\alpha^{9}\beta^{4} + 12\alpha^{8}\beta^{5} + 14\alpha^{7}\beta^{6} + 12\alpha^{6}\beta^{7} + 10\alpha^{5}\beta^{8} + 4\alpha^{4}\beta^{9} + 2\alpha^{3}\beta^{10} + 2\alpha^{2}\beta^{11}\end{pmatrix} \\
\hat{101000} = &\begin{pmatrix} 2\alpha^{11}\beta^{2} + 2\alpha^{10}\beta^{3} + 4\alpha^{9}\beta^{4} + 10\alpha^{8}\beta^{5} + 12\alpha^{7}\beta^{6} + 14\alpha^{6}\beta^{7} + 12\alpha^{5}\beta^{8} + 6\alpha^{4}\beta^{9} + 2\alpha^{3}\beta^{10} \\ 2\alpha^{10}\beta^{3} + 6\alpha^{9}\beta^{4} + 12\alpha^{8}\beta^{5} + 14\alpha^{7}\beta^{6} + 12\alpha^{6}\beta^{7} + 10\alpha^{5}\beta^{8} + 4\alpha^{4}\beta^{9} + 2\alpha^{3}\beta^{10} + 2\alpha^{2}\beta^{11}\end{pmatrix} \\
\hat{101001} = &\begin{pmatrix} 3\alpha^{10}\beta^{3} + 6\alpha^{9}\beta^{4} + 10\alpha^{8}\beta^{5} + 14\alpha^{7}\beta^{6} + 12\alpha^{6}\beta^{7} + 10\alpha^{5}\beta^{8} + 6\alpha^{4}\beta^{9} + 2\alpha^{3}\beta^{10} + \alpha^{2}\beta^{11} \\ \alpha^{11}\beta^{2} + \alpha^{10}\beta^{3} + 5\alpha^{9}\beta^{4} + 12\alpha^{8}\beta^{5} + 14\alpha^{7}\beta^{6} + 14\alpha^{6}\beta^{7} + 10\alpha^{5}\beta^{8} + 4\alpha^{4}\beta^{9} + \alpha^{3}\beta^{10} + \alpha^{2}\beta^{11} + \alpha\beta^{12}\end{pmatrix} \\
\hat{101010} = &\begin{pmatrix} 2\alpha^{10}\beta^{3} + 10\alpha^{9}\beta^{4} + 10\alpha^{8}\beta^{5} + 6\alpha^{7}\beta^{6} + 14\alpha^{6}\beta^{7} + 14\alpha^{5}\beta^{8} + 6\alpha^{4}\beta^{9} + 2\alpha^{3}\beta^{10} \\ 4\alpha^{10}\beta^{3} + 6\alpha^{9}\beta^{4} + 6\alpha^{8}\beta^{5} + 14\alpha^{7}\beta^{6} + 18\alpha^{6}\beta^{7} + 10\alpha^{5}\beta^{8} + 2\alpha^{4}\beta^{9} + 2\alpha^{3}\beta^{10} + 2\alpha^{2}\beta^{11}\end{pmatrix} \\
\hat{101011} = &\begin{pmatrix} 4\alpha^{10}\beta^{3} + 6\alpha^{9}\beta^{4} + 8\alpha^{8}\beta^{5} + 14\alpha^{7}\beta^{6} + 12\alpha^{6}\beta^{7} + 10\alpha^{5}\beta^{8} + 8\alpha^{4}\beta^{9} + 2\alpha^{3}\beta^{10} \\ \alpha^{10}\beta^{3} + 7\alpha^{9}\beta^{4} + 12\alpha^{8}\beta^{5} + 14\alpha^{7}\beta^{6} + 14\alpha^{6}\beta^{7} + 8\alpha^{5}\beta^{8} + 4\alpha^{4}\beta^{9} + 2\alpha^{3}\beta^{10} + \alpha^{2}\beta^{11} + \alpha\beta^{12}\end{pmatrix} \\
\end{aligned}
\end{equation}
\begin{equation}
\begin{aligned}
\hat{101100} = &\begin{pmatrix} 3\alpha^{10}\beta^{3} + 6\alpha^{9}\beta^{4} + 10\alpha^{8}\beta^{5} + 14\alpha^{7}\beta^{6} + 12\alpha^{6}\beta^{7} + 10\alpha^{5}\beta^{8} + 6\alpha^{4}\beta^{9} + 2\alpha^{3}\beta^{10} + \alpha^{2}\beta^{11} \\ \alpha^{11}\beta^{2} + \alpha^{10}\beta^{3} + 5\alpha^{9}\beta^{4} + 12\alpha^{8}\beta^{5} + 14\alpha^{7}\beta^{6} + 14\alpha^{6}\beta^{7} + 10\alpha^{5}\beta^{8} + 4\alpha^{4}\beta^{9} + \alpha^{3}\beta^{10} + \alpha^{2}\beta^{11} + \alpha\beta^{12}\end{pmatrix} \\
\hat{101101} = &\begin{pmatrix} \alpha^{11}\beta^{2} + 6\alpha^{9}\beta^{4} + 16\alpha^{8}\beta^{5} + 12\alpha^{7}\beta^{6} + 8\alpha^{6}\beta^{7} + 10\alpha^{5}\beta^{8} + 8\alpha^{4}\beta^{9} + 3\alpha^{3}\beta^{10} \\ 2\alpha^{10}\beta^{3} + 4\alpha^{9}\beta^{4} + 11\alpha^{8}\beta^{5} + 20\alpha^{7}\beta^{6} + 14\alpha^{6}\beta^{7} + 4\alpha^{5}\beta^{8} + 4\alpha^{4}\beta^{9} + 4\alpha^{3}\beta^{10} + \beta^{13}\end{pmatrix} \\
\hat{101110} = &\begin{pmatrix} 4\alpha^{10}\beta^{3} + 6\alpha^{9}\beta^{4} + 8\alpha^{8}\beta^{5} + 14\alpha^{7}\beta^{6} + 12\alpha^{6}\beta^{7} + 10\alpha^{5}\beta^{8} + 8\alpha^{4}\beta^{9} + 2\alpha^{3}\beta^{10} \\ \alpha^{10}\beta^{3} + 7\alpha^{9}\beta^{4} + 12\alpha^{8}\beta^{5} + 14\alpha^{7}\beta^{6} + 14\alpha^{6}\beta^{7} + 8\alpha^{5}\beta^{8} + 4\alpha^{4}\beta^{9} + 2\alpha^{3}\beta^{10} + \alpha^{2}\beta^{11} + \alpha\beta^{12}\end{pmatrix} \\
\hat{101111} = &\begin{pmatrix} \alpha^{10}\beta^{3} + 8\alpha^{9}\beta^{4} + 12\alpha^{8}\beta^{5} + 12\alpha^{7}\beta^{6} + 14\alpha^{6}\beta^{7} + 8\alpha^{5}\beta^{8} + 4\alpha^{4}\beta^{9} + 4\alpha^{3}\beta^{10} + \alpha^{2}\beta^{11} \\ 2\alpha^{10}\beta^{3} + 5\alpha^{9}\beta^{4} + 12\alpha^{8}\beta^{5} + 16\alpha^{7}\beta^{6} + 12\alpha^{6}\beta^{7} + 10\alpha^{5}\beta^{8} + 4\alpha^{4}\beta^{9} + 2\alpha^{2}\beta^{11} + \alpha\beta^{12}\end{pmatrix} \\
\hat{110000} = &\begin{pmatrix} \alpha^{12}\beta + \alpha^{11}\beta^{2} + 2\alpha^{10}\beta^{3} + 4\alpha^{9}\beta^{4} + 8\alpha^{8}\beta^{5} + 14\alpha^{7}\beta^{6} + 14\alpha^{6}\beta^{7} + 12\alpha^{5}\beta^{8} + 7\alpha^{4}\beta^{9} + \alpha^{3}\beta^{10} \\ 2\alpha^{10}\beta^{3} + 8\alpha^{9}\beta^{4} + 10\alpha^{8}\beta^{5} + 12\alpha^{7}\beta^{6} + 14\alpha^{6}\beta^{7} + 8\alpha^{5}\beta^{8} + 6\alpha^{4}\beta^{9} + 4\alpha^{3}\beta^{10}\end{pmatrix} \\
\hat{110001} = &\begin{pmatrix} \alpha^{11}\beta^{2} + 2\alpha^{10}\beta^{3} + 6\alpha^{9}\beta^{4} + 10\alpha^{8}\beta^{5} + 12\alpha^{7}\beta^{6} + 14\alpha^{6}\beta^{7} + 10\alpha^{5}\beta^{8} + 6\alpha^{4}\beta^{9} + 3\alpha^{3}\beta^{10} \\ 3\alpha^{10}\beta^{3} + 6\alpha^{9}\beta^{4} + 10\alpha^{8}\beta^{5} + 14\alpha^{7}\beta^{6} + 12\alpha^{6}\beta^{7} + 10\alpha^{5}\beta^{8} + 6\alpha^{4}\beta^{9} + 2\alpha^{3}\beta^{10} + \alpha^{2}\beta^{11}\end{pmatrix} \\
\hat{110010} = &\begin{pmatrix} \alpha^{11}\beta^{2} + 4\alpha^{10}\beta^{3} + 4\alpha^{9}\beta^{4} + 8\alpha^{8}\beta^{5} + 14\alpha^{7}\beta^{6} + 12\alpha^{6}\beta^{7} + 12\alpha^{5}\beta^{8} + 8\alpha^{4}\beta^{9} + \alpha^{3}\beta^{10} \\ \alpha^{11}\beta^{2} + 2\alpha^{10}\beta^{3} + 6\alpha^{9}\beta^{4} + 10\alpha^{8}\beta^{5} + 12\alpha^{7}\beta^{6} + 14\alpha^{6}\beta^{7} + 10\alpha^{5}\beta^{8} + 6\alpha^{4}\beta^{9} + 3\alpha^{3}\beta^{10}\end{pmatrix} \\
\hat{110011} = &\begin{pmatrix} \alpha^{11}\beta^{2} + 2\alpha^{10}\beta^{3} + 8\alpha^{9}\beta^{4} + 10\alpha^{8}\beta^{5} + 6\alpha^{7}\beta^{6} + 14\alpha^{6}\beta^{7} + 16\alpha^{5}\beta^{8} + 6\alpha^{4}\beta^{9} + \alpha^{3}\beta^{10} \\ \alpha^{10}\beta^{3} + 6\alpha^{9}\beta^{4} + 16\alpha^{8}\beta^{5} + 14\alpha^{7}\beta^{6} + 6\alpha^{6}\beta^{7} + 10\alpha^{5}\beta^{8} + 8\alpha^{4}\beta^{9} + 2\alpha^{3}\beta^{10} + \alpha^{2}\beta^{11}\end{pmatrix} \\
\hat{110100} = &\begin{pmatrix} \alpha^{11}\beta^{2} + 3\alpha^{10}\beta^{3} + 6\alpha^{9}\beta^{4} + 6\alpha^{8}\beta^{5} + 12\alpha^{7}\beta^{6} + 20\alpha^{6}\beta^{7} + 10\alpha^{5}\beta^{8} + 2\alpha^{4}\beta^{9} + 3\alpha^{3}\beta^{10} + \alpha^{2}\beta^{11} \\ \alpha^{11}\beta^{2} + 3\alpha^{10}\beta^{3} + 2\alpha^{9}\beta^{4} + 10\alpha^{8}\beta^{5} + 20\alpha^{7}\beta^{6} + 12\alpha^{6}\beta^{7} + 6\alpha^{5}\beta^{8} + 6\alpha^{4}\beta^{9} + 3\alpha^{3}\beta^{10} + \alpha^{2}\beta^{11}\end{pmatrix} \\
\hat{110101} = &\begin{pmatrix} 4\alpha^{10}\beta^{3} + 6\alpha^{9}\beta^{4} + 8\alpha^{8}\beta^{5} + 14\alpha^{7}\beta^{6} + 12\alpha^{6}\beta^{7} + 10\alpha^{5}\beta^{8} + 8\alpha^{4}\beta^{9} + 2\alpha^{3}\beta^{10} \\ \alpha^{10}\beta^{3} + 7\alpha^{9}\beta^{4} + 12\alpha^{8}\beta^{5} + 14\alpha^{7}\beta^{6} + 14\alpha^{6}\beta^{7} + 8\alpha^{5}\beta^{8} + 4\alpha^{4}\beta^{9} + 2\alpha^{3}\beta^{10} + \alpha^{2}\beta^{11} + \alpha\beta^{12}\end{pmatrix} \\
\hat{110110} = &\begin{pmatrix} 2\alpha^{11}\beta^{2} + 2\alpha^{10}\beta^{3} + 4\alpha^{9}\beta^{4} + 10\alpha^{8}\beta^{5} + 12\alpha^{7}\beta^{6} + 14\alpha^{6}\beta^{7} + 12\alpha^{5}\beta^{8} + 6\alpha^{4}\beta^{9} + 2\alpha^{3}\beta^{10} \\ 2\alpha^{10}\beta^{3} + 6\alpha^{9}\beta^{4} + 12\alpha^{8}\beta^{5} + 14\alpha^{7}\beta^{6} + 12\alpha^{6}\beta^{7} + 10\alpha^{5}\beta^{8} + 4\alpha^{4}\beta^{9} + 2\alpha^{3}\beta^{10} + 2\alpha^{2}\beta^{11}\end{pmatrix} \\
\hat{110111} = &\begin{pmatrix} 3\alpha^{10}\beta^{3} + 6\alpha^{9}\beta^{4} + 10\alpha^{8}\beta^{5} + 14\alpha^{7}\beta^{6} + 12\alpha^{6}\beta^{7} + 10\alpha^{5}\beta^{8} + 6\alpha^{4}\beta^{9} + 2\alpha^{3}\beta^{10} + \alpha^{2}\beta^{11} \\ \alpha^{10}\beta^{3} + 8\alpha^{9}\beta^{4} + 12\alpha^{8}\beta^{5} + 12\alpha^{7}\beta^{6} + 14\alpha^{6}\beta^{7} + 8\alpha^{5}\beta^{8} + 4\alpha^{4}\beta^{9} + 4\alpha^{3}\beta^{10} + \alpha^{2}\beta^{11}\end{pmatrix} \\
\hat{111000} = &\begin{pmatrix} 2\alpha^{10}\beta^{3} + 6\alpha^{9}\beta^{4} + 14\alpha^{8}\beta^{5} + 14\alpha^{7}\beta^{6} + 6\alpha^{6}\beta^{7} + 10\alpha^{5}\beta^{8} + 10\alpha^{4}\beta^{9} + 2\alpha^{3}\beta^{10} \\ 2\alpha^{11}\beta^{2} + 2\alpha^{10}\beta^{3} + 2\alpha^{9}\beta^{4} + 10\alpha^{8}\beta^{5} + 18\alpha^{7}\beta^{6} + 14\alpha^{6}\beta^{7} + 6\alpha^{5}\beta^{8} + 6\alpha^{4}\beta^{9} + 4\alpha^{3}\beta^{10}\end{pmatrix} \\
\hat{111001} = &\begin{pmatrix} \alpha^{11}\beta^{2} + \alpha^{10}\beta^{3} + 6\alpha^{9}\beta^{4} + 12\alpha^{8}\beta^{5} + 12\alpha^{7}\beta^{6} + 14\alpha^{6}\beta^{7} + 10\alpha^{5}\beta^{8} + 4\alpha^{4}\beta^{9} + 3\alpha^{3}\beta^{10} + \alpha^{2}\beta^{11} \\ 2\alpha^{10}\beta^{3} + 6\alpha^{9}\beta^{4} + 12\alpha^{8}\beta^{5} + 14\alpha^{7}\beta^{6} + 12\alpha^{6}\beta^{7} + 10\alpha^{5}\beta^{8} + 4\alpha^{4}\beta^{9} + 2\alpha^{3}\beta^{10} + 2\alpha^{2}\beta^{11}\end{pmatrix} \\
\hat{111010} = &\begin{pmatrix} 4\alpha^{10}\beta^{3} + 6\alpha^{9}\beta^{4} + 8\alpha^{8}\beta^{5} + 14\alpha^{7}\beta^{6} + 12\alpha^{6}\beta^{7} + 10\alpha^{5}\beta^{8} + 8\alpha^{4}\beta^{9} + 2\alpha^{3}\beta^{10} \\ 2\alpha^{10}\beta^{3} + 8\alpha^{9}\beta^{4} + 10\alpha^{8}\beta^{5} + 12\alpha^{7}\beta^{6} + 14\alpha^{6}\beta^{7} + 8\alpha^{5}\beta^{8} + 6\alpha^{4}\beta^{9} + 4\alpha^{3}\beta^{10}\end{pmatrix} \\
\hat{111011} = &\begin{pmatrix} \alpha^{10}\beta^{3} + 8\alpha^{9}\beta^{4} + 12\alpha^{8}\beta^{5} + 12\alpha^{7}\beta^{6} + 14\alpha^{6}\beta^{7} + 8\alpha^{5}\beta^{8} + 4\alpha^{4}\beta^{9} + 4\alpha^{3}\beta^{10} + \alpha^{2}\beta^{11} \\ 3\alpha^{10}\beta^{3} + 6\alpha^{9}\beta^{4} + 10\alpha^{8}\beta^{5} + 14\alpha^{7}\beta^{6} + 12\alpha^{6}\beta^{7} + 10\alpha^{5}\beta^{8} + 6\alpha^{4}\beta^{9} + 2\alpha^{3}\beta^{10} + \alpha^{2}\beta^{11}\end{pmatrix} \\
\hat{111100} = &\begin{pmatrix} \alpha^{11}\beta^{2} + \alpha^{10}\beta^{3} + 6\alpha^{9}\beta^{4} + 12\alpha^{8}\beta^{5} + 12\alpha^{7}\beta^{6} + 14\alpha^{6}\beta^{7} + 10\alpha^{5}\beta^{8} + 4\alpha^{4}\beta^{9} + 3\alpha^{3}\beta^{10} + \alpha^{2}\beta^{11} \\ 2\alpha^{10}\beta^{3} + 6\alpha^{9}\beta^{4} + 12\alpha^{8}\beta^{5} + 14\alpha^{7}\beta^{6} + 12\alpha^{6}\beta^{7} + 10\alpha^{5}\beta^{8} + 4\alpha^{4}\beta^{9} + 2\alpha^{3}\beta^{10} + 2\alpha^{2}\beta^{11}\end{pmatrix} \\
\hat{111101} = &\begin{pmatrix} 2\alpha^{10}\beta^{3} + 5\alpha^{9}\beta^{4} + 12\alpha^{8}\beta^{5} + 16\alpha^{7}\beta^{6} + 12\alpha^{6}\beta^{7} + 10\alpha^{5}\beta^{8} + 4\alpha^{4}\beta^{9} + 2\alpha^{2}\beta^{11} + \alpha\beta^{12} \\ \alpha^{10}\beta^{3} + 8\alpha^{9}\beta^{4} + 12\alpha^{8}\beta^{5} + 12\alpha^{7}\beta^{6} + 14\alpha^{6}\beta^{7} + 8\alpha^{5}\beta^{8} + 4\alpha^{4}\beta^{9} + 4\alpha^{3}\beta^{10} + \alpha^{2}\beta^{11}\end{pmatrix} \\
\hat{111110} = &\begin{pmatrix} \alpha^{10}\beta^{3} + 8\alpha^{9}\beta^{4} + 12\alpha^{8}\beta^{5} + 12\alpha^{7}\beta^{6} + 14\alpha^{6}\beta^{7} + 8\alpha^{5}\beta^{8} + 4\alpha^{4}\beta^{9} + 4\alpha^{3}\beta^{10} + \alpha^{2}\beta^{11} \\ 3\alpha^{10}\beta^{3} + 6\alpha^{9}\beta^{4} + 10\alpha^{8}\beta^{5} + 14\alpha^{7}\beta^{6} + 12\alpha^{6}\beta^{7} + 10\alpha^{5}\beta^{8} + 6\alpha^{4}\beta^{9} + 2\alpha^{3}\beta^{10} + \alpha^{2}\beta^{11}\end{pmatrix} \\
\hat{111111} = &\begin{pmatrix} 3\alpha^{10}\beta^{3} + 4\alpha^{9}\beta^{4} + 10\alpha^{8}\beta^{5} + 20\alpha^{7}\beta^{6} + 12\alpha^{6}\beta^{7} + 4\alpha^{5}\beta^{8} + 6\alpha^{4}\beta^{9} + 4\alpha^{3}\beta^{10} + \alpha^{2}\beta^{11} \\ 7\alpha^{9}\beta^{4} + 16\alpha^{8}\beta^{5} + 14\alpha^{7}\beta^{6} + 8\alpha^{6}\beta^{7} + 8\alpha^{5}\beta^{8} + 8\alpha^{4}\beta^{9} + 2\alpha^{3}\beta^{10} + \alpha\beta^{12}\end{pmatrix} \\
\end{aligned}
\label{eq:d3}
\end{equation}

\clearpage
\section{Appendix C - Specific states produced via transversal injection.}

In this section we give an example of a specific solution for each of the expressions in Eq. \ref{eq:d3}, for the $d=3$ codes where an $|A\rangle$ state, suitable to implement the $T$-gate exists, for an initial state, written in polar form, $(|\chi\rangle)^{\otimes 13}=\left(\cos\left(\frac{\theta}{2}\right)|0\rangle+e^{i\phi}\sin\left(\frac{\theta}{2}\right)|1\rangle\right)^{\otimes 13}$, where $\left(\theta = 2.44580563149781, \phi = 1.3616970885685595\right)$.  It should be noted that this solution is {\em not unique}. \\
\\
The trivial $X$-trajectories and $Z$-trajectories corresponding to the binary vectors, $\hat{010011}$ and $\hat{010110}$ produce the state  $\left(|0\rangle_L + e^{-i\pi/4}|1\rangle_L\right)/\sqrt{2}$, while the $Z$-trajectories corresponding to the binary vectors, $\hat{011010}$ and $\hat{110010}$ produce the state $\left(|0\rangle_L + e^{i\pi/4}|1\rangle_L\right)/\sqrt{2}$.  Both of these states can be used to realise a logical $T$-gate.  All 64 states produced for a $d=3$ code with the input above are illustrated in Fig. \ref{fig:d3}.  
\\
\\
It is the focus of future work to investigate the states produced by transversal injection and how these states can be utilised even though their preparation is probabilistic, but heralded.
\\
\twocolumngrid
\begin{tabular}{ |p{1.2cm}||p{3.6cm}|p{3.6cm}|  }
 \hline
 $Z$-traj & $\theta_L$ & $\phi_L$ \\
 \hline
$\hat{000000}$ & 1.477251531844677 & -2.286354913393752 \\
$\hat{000001}$ & 2.4598774702571236 & -2.020056925558729 \\
$\hat{000010}$ & 2.014303116800432 & -0.37430535389058534 \\
$\hat{000011}$ & 2.0000126349446785 & 1.0875192568311256 \\
$\hat{000100}$ & 2.4598774702571236 & -2.020056925558729 \\
$\hat{000101}$ & 2.2207374531113 & 2.2914520972008927 \\
$\hat{000110}$ & 2.0000126349446785 & 1.0875192568311256 \\
$\hat{000111}$ & 1.5899082238746929 & -0.7176718165630172 \\
$\hat{001000}$ & 0.6817151833326703 & 2.020056925558729 \\
$\hat{001001}$ & 2.6105434793012248 & 2.3851586668749465 \\
$\hat{001010}$ & 0.6010286049447089 & 2.066850378564112 \\
$\hat{001011}$ & 1.505967601101266 & 0.5553720557783275 \\
$\hat{001100}$ & 2.3652945274675323 & -1.1917847714989986 \\
$\hat{001101}$ & 0.4526355842480339 & -2.1740432175058464 \\
$\hat{001110}$ & 2.0155964891204774 & -1.1347363564914656 \\
$\hat{001111}$ & 2.0275033715124824 & 0.5744495240028695 \\
$\hat{010000}$ & 1.127289536789362 & 0.37430535389058534 \\
$\hat{010001}$ & 0.6010286049447089 & 2.066850378564112 \\
$\hat{010010}$ & 1.113903157150333 & 1.5597871287260436 \\
$\hat{{\bf 010011}}$ & {\bf 1.5707963267679674} & {\bf -0.7853981633911511} \\
$\hat{010100}$ & 0.6010286049447089 & 2.066850378564112 \\
$\hat{010101}$ & 0.8387195533772056 & -2.58989408067244 \\
$\hat{{\bf 010110}}$ & {\bf 1.5707963267679674} & {\bf -0.7853981633911511} \\
$\hat{010111}$ & 1.791506609499127 & 0.6159885397597338 \\
$\hat{011000}$ & 2.0000126349446785 & 1.0875192568311256 \\
$\hat{011001}$ & 1.6356250524885274 & -0.5553720557783275 \\
$\hat{\bf{011010}}$ & {\bf 1.5707963268218257} & {\bf 0.7853981633911511} \\
$\hat{011011}$ & 2.2207374531113 & 2.2914520972008927 \\
$\hat{011100}$ & 1.125996164469316 & 1.1347363564914656 \\
$\hat{011101}$ & 2.316857739052819 & -2.723267254060598 \\
$\hat{011110}$ & 1.375057538833144 & 1.7128079495289403 \\
$\hat{011111}$ & 2.2924864906921796 & 0.023764711852680698 \\

 \hline
\end{tabular}

\begin{tabular}{ |p{1.2cm}||p{3.6cm}|p{3.6cm}|  }
 \hline
 $Z$-traj & $\theta_L$ & $\phi_L$ \\
 \hline

$\hat{100000}$ & 0.6817151833326703 & 2.020056925558729 \\
$\hat{100001}$ & 2.3652945274675323 & -1.1917847714989986 \\
$\hat{100010}$ & 0.6010286049447089 & 2.066850378564112 \\
$\hat{100011}$ & 2.0155964891204774 & -1.1347363564914656 \\
$\hat{100100}$ & 2.6105434793012248 & 2.3851586668749465 \\
$\hat{100101}$ & 0.4526355842480339 & -2.1740432175058464 \\
$\hat{100110}$ & 1.505967601101266 & 0.5553720557783275 \\
$\hat{100111}$ & 2.0275033715124824 & 0.5744495240028695 \\
$\hat{101000}$ & 2.2207374531113 & 2.2914520972008927 \\
$\hat{101001}$ & 2.68895706934176 & 2.174043217505847 \\
$\hat{101010}$ & 2.3028731002125875 & 2.58989408067244 \\
$\hat{101011}$ & 2.316857739052819 & -2.723267254060598 \\
$\hat{101100}$ & 2.68895706934176 & 2.174043217505847 \\
$\hat{101101}$ & 2.760788786089281 & -1.1274336283037671 \\
$\hat{101110}$ & 2.316857739052819 & -2.723267254060598 \\
$\hat{101111}$ & 2.370102532312507 & 1.9493577457230364 \\
$\hat{110000}$ & 2.0000126349446785 & 1.0875192568311256 \\
$\hat{110001}$ & 1.125996164469316 & 1.1347363564914656 \\
$\hat{{\bf 110010}}$ & {\bf 1.5707963268218257} & {\bf 0.7853981633911511} \\
$\hat{110011}$ & 1.375057538833144 & 1.7128079495289403 \\
$\hat{110100}$ & 1.6356250524885274 & -0.5553720557783275 \\
$\hat{110101}$ & 2.316857739052819 & -2.723267254060598 \\
$\hat{110110}$ & 2.2207374531113 & 2.2914520972008927 \\
$\hat{110111}$ & 2.2924864906921796 & 0.023764711852680698 \\
$\hat{111000}$ & 1.5516844297151002 & 0.717671816563017 \\
$\hat{111001}$ & 2.0275033715124824 & 0.5744495240028695 \\
$\hat{111010}$ & 1.791506609499127 & 0.6159885397597338 \\
$\hat{111011}$ & 0.8491061628976135 & -0.023764711852680698 \\
$\hat{111100}$ & 2.0275033715124824 & 0.5744495240028695 \\
$\hat{111101}$ & 0.7714901212772858 & -1.9493577457230364 \\
$\hat{111110}$ & 0.8491061628976135 & -0.023764711852680698 \\
$\hat{111111}$ & 1.7094217253738777 & 2.868327212586131\\
\hline
\end{tabular}

\end{document}